\documentclass[12pt]{article}

\usepackage[T1]{fontenc}
\usepackage{lmodern}

\usepackage{amssymb,latexsym}
\usepackage{microtype}
\usepackage{natbib}
\usepackage[unicode=true, bookmarks=true, bookmarksnumbered=false, bookmarksopen=false, breaklinks=true, pdfborder={0 0 0},backref=false]{hyperref}

\usepackage[dvipsnames]{xcolor}
\hypersetup{
    colorlinks,
    linkcolor={red!80!black},
    citecolor={blue!60!black},
    urlcolor={green!35!black},
}

\usepackage{amsmath}
\usepackage{amsthm}

\DeclareFontFamily{U}{mathx}{}
\DeclareFontShape{U}{mathx}{m}{n}{<-> mathx10}{}
\DeclareSymbolFont{mathx}{U}{mathx}{m}{n}
\DeclareMathAccent{\widehat}{0}{mathx}{"70}
\DeclareMathAccent{\widecheck}{0}{mathx}{"71}

\usepackage{graphicx}
\usepackage{cases}
\usepackage{float}
\usepackage{subcaption}
\usepackage{multirow}
\usepackage{enumitem}
\setlist[itemize]{itemsep=0pt, topsep=3pt}

\DeclareMathAlphabet{\mathmybb}{U}{bbold}{m}{n}

\usepackage{tikz}
\usetikzlibrary{shapes,positioning,intersections,quotes,patterns}
\usetikzlibrary{decorations.pathreplacing}

\DeclareFontFamily{U}{mathx}{\hyphenchar\font45}
\DeclareFontShape{U}{mathx}{m}{n}{
      <5> <6> <7> <8> <9> <10>
      <10.95> <12> <14.4> <17.28> <20.74> <24.88>
      mathx10
      }{}
\DeclareSymbolFont{mathx}{U}{mathx}{m}{n}
\DeclareFontSubstitution{U}{mathx}{m}{n}
\DeclareMathAccent{\widebar}{0}{mathx}{"73}

\usepackage{geometry,setspace}
\allowdisplaybreaks

\theoremstyle{plain}

\newtheorem{lemma}{Lemma}
\newtheorem{proposition}{Proposition}

\theoremstyle{definition}

\newtheorem{definition}{Definition}
\newtheorem{remark}{Remark}
\newtheorem{examplex}{Example}
\newenvironment{example}
  {\pushQED{\qed}\examplex}
  {\popQED\endexamplex}

\def\A{\mathcal{A}}
\def\G{\mathcal{G}}

\def\E{\mathcal{E}}

\newcommand*{\CE}{\mathit{CE}}

\newcommand*{\supp}{\mathrm{supp}}

\DeclareMathOperator*{\argmax}{arg\,max}

\let\widebar\relax
\usepackage{newtxtext}
\usepackage{newtxmath}

\usepackage{titlesec}

\titleformat{\section}
  {\normalfont\large\bfseries}
  {\thesection}{1em}{}

\titlespacing*{\section}
  {0pt}{2.5ex plus 1ex minus .2ex}{1.5ex plus .2ex}

\titleformat{\subsection}
  {\normalfont\normalsize\bfseries}
  {\thesubsection}{1em}{}

\titlespacing*{\subsection}
  {0pt}{2ex plus .8ex minus .2ex}{1ex plus .2ex}

\begin{document}
\title{Preference Evolution under Partner Choice\footnote{We thank Ingela Alger, Yi-Chun Chen, Jeffrey Ely, Wei He, Gaoji Hu, Luis Izquierdo, Segismundo Izquierdo, Sam Jindani, Jinwoo Kim, Ce Liu, Qingmin Liu, Zhuoran Lu, Weicheng Min, Jonathan Newton, Fanqi Shi, Xiang Sun, Yifei Sun, Ina Taneva, Qianfeng Tang, Yi Tong, Matthijs van Veelen, J\"{o}rgen Weibull, Yiqing Xing, Nate Yoder, Siyang Xiong, Hanzhe Zhang, Kun Zhang, and Jidong Zhou for constructive comments and suggestions. We also thank seminar and conference participants at Renmin University of China, Peking University, Capital University of Economics and Business, Shanghai Jiao Tong University, UC Riverside, NASM 2024 (Vanderbilt), the 7th World Congress of the Game Theory Society (PKU), the 2024 HUST Economic Theory Workshop, the LEG Online Seminar, CETC 2025 (Montr\'{e}al), CoED 2025 (Essex), ESWC 2025 (Seoul), and the 2026 Summer Workshop on Microeconomic Theory (CUFE) for valuable discussions. Rongkai Liu provided helpful research assistance. Wang acknowledges financial support from the National Natural Science Foundation of China (No.~72403188). The usual disclaimer applies.}}

\author{Ziwei Wang\footnote{Guanghua School of Management, Peking University. Email: zwang.econ@gmail.com.} \and Jiabin Wu\footnote{Department of Economics, University of Oregon. Email: jwu5@uoregon.edu.}}

\date{\vspace{0.15in} August 16, 2026}
\maketitle

\begin{abstract}

\noindent 
We study preference evolution when agents choose whom to interact with. In the short run, subjective preferences influence both partner choice and strategic behavior, which affect their material payoffs. These payoffs, in turn, determine how preferences evolve in the long run. To model this ``match-to-interact'' process, we combine stable matching with correlated equilibrium, allowing matched agents to coordinate on self-enforcing agreements. We show that evolution under endogenous matching need not favor either purely selfish or purely efficient preferences. Instead, we identify two classes of preferences that combine these concerns nonlinearly and can prevail because they sustain efficient play while protecting agents from exploitation. Under incomplete information, however, only one of these classes retains its evolutionary advantage. This class combines efficiency with parochialism, a form of group-level selfishness that strengthens incentives for self-sorting.
\vspace{0.1in}

\end{abstract}

\renewcommand*{\thefootnote}{\arabic{footnote}}
\setcounter{footnote}{0}




\newpage

\section{Introduction}
Economic analysis typically treats decision makers' preferences as fixed and exogenous. Preferences, however, may themselves be the product of a lengthy evolutionary process. This raises a fundamental question: why do some preferences persist while others disappear? \citet{GuthYaari1992} and \citet{Guth1995} introduce the ``indirect evolutionary approach,'' which provides a useful theoretical framework for understanding preference evolution: preferences shape behavior, behavior determines fitness, and fitness, in turn, governs the evolution of preferences.\footnote{Earlier contributions developing related ideas include \citet{Becker1976}, \citet{Hirshleifer1977}, \citet{RubinandPaul1979}, and \citet{Frank1987}.} A preference type is considered evolutionarily stable if, when predominant in a population, it can resist invasion by alternative types; see recent surveys by \citet{AlgerWeibull2019ARE} and \citet{Alger2022}.\footnote{See also \citet{RobsonandSamuelson2011} for a critical assessment of this approach.}

Most work in this tradition studies behavior in two-player games among individuals paired through some exogenous random matching process.\footnote{A few papers instead consider multilateral games \citep{Lehmann2015Evolution, AlgerAndWeibull2016GEB, Algeretal2020JET} or environments in which all individuals ``play the field'' \citep{Lahkar2019JME, bandhu2023survival}. Matching remains exogenous in the former and plays no role in the latter.} This focus, however, omits another important behavioral margin. Preferences may shape not only how individuals play after being matched, but also whom they choose to interact with and, hence, how matches form in the first place. When preferences determine both strategic behavior and partner choice, matching itself becomes endogenous in the evolutionary process.

This paper develops a model of preference evolution with endogenous matching that accommodates a rich space of preference types, any finite symmetric two-player game, and potentially incomplete information about other agents' preference types. We model the joint determination of matching and play in a continuum population by combining the noncooperative concept of correlated equilibrium and the cooperative notion of stable matching.\footnote{Formally, our framework is a large matching market with contracts. We draw on recent results in this literature to establish the existence of stable outcomes in our setting \citep{CheKimKojima2019, Greinecker2021, Jagadeesan2024, carmona2024}.} Previous studies by \citet{JacksonWatts2010}, \citet{Xing2020}, and \citet{Garrido-LucerolLaraki2021WP} adopt a similar approach but rely on Nash equilibrium (or subgame-perfect Nash equilibrium) to discipline play within matches. The additional coordination permitted by correlated equilibrium is economically important when efficient interaction requires partners to take different roles. An agreement can then randomize over role assignments and thereby produce a more balanced division of expected payoff. Stability, in turn, captures the idea that unsatisfied individuals can communicate and jointly deviate by forming a new match and coordinating on an agreement that benefits both. A stable outcome can therefore be viewed as the reduced-form outcome of a frictionless process of partner formation.

We begin with a complete-information benchmark in which preference types are observable. To characterize stable outcomes in this environment, we propose a concept called \textit{complete-information stability} for a continuum population. Specifically, it requires that every matched pair of individuals plays a correlated equilibrium, termed \textit{internal stability}, and that no pair of agents can coordinate on a correlated equilibrium that makes both strictly better off, termed \textit{external stability}. Given a population composed of heterogeneous preference types, agents' preferences jointly shape both whom they match with and how they play within a match, and thereby the material payoffs they obtain in a stable outcome. We then evaluate the evolutionary performance of preference types by comparing their average material payoffs across stable outcomes.

We show that two natural benchmarks---selfish and efficient preferences---can each fail to prevail in evolution. Selfish preferences provide an inherent safeguard against exploitation because selfish agents always seek to maximize their own material payoffs. Yet they may sustain inefficient equilibrium play, leaving selfish agents at an evolutionary disadvantage.\footnote{More generally, altruistic preferences that are not aligned with joint material payoff may suffer from the same weakness. This contrasts with the conventional view in theoretical biology, dating back to \citet{Hamilton1964a,Hamilton1964b}, that assortative interaction---whether arising through kinship, reciprocity, or group selection---can sustain altruism. With endogenous partner choice, however, a competing type can itself generate positive assortative matching, turning inefficient play by altruistic agents into an evolutionary disadvantage.} Efficient preferences, by contrast, align agents' incentives with joint material payoff, which ensures that efficient play can be sustained in equilibrium. Their weakness is excessive benevolence: efficient agents may accept an unfavorable division of the joint material payoff and thus become vulnerable to exploitation.\footnote{A preference for efficiency admits two interpretations. On the one hand, it can be viewed as a ``disinterested'' preference, whereby an individual is solely focused on the social objective of maximizing the total material payoff. This perspective is in line with \textit{utilitarianism}: conditional on her partner's behavior, the individual regards an action as morally right when it maximizes the pair's welfare \citep{Mill1863}. On the other hand, the preference for efficiency can also be seen as a form of altruism, whereby an individual places equal importance on her own material payoff and that of her opponent.}

These weaknesses can be overcome by combining selfishness and efficiency in nonlinear ways. We identify two classes of preference types that do so and can thus prevail in evolution. The first consists of \textit{morally constrained selfish} types, whose preferences rank the two objectives hierarchically. Agents with such preferences place first priority on efficient play with their partners and then, conditional on efficiency, maximize their own material payoffs. The efficiency objective disciplines behavior because, whenever agents of this type play inefficiently, two of them can profitably rematch and implement an efficient equilibrium. The selfish component, in turn, protects agents from unfavorable payoff divisions: agents who receive too little can rematch and use a symmetrized efficient equilibrium to divide the maximum joint material payoff equally. As a remarkable consequence of endogenous partner choice, agents with preferences in this class can achieve both efficiency in payoff realization and fairness in distribution.

The second class comprises \textit{parochial efficient} types, whose preferences also embody a hierarchical ordering of objectives. Agents with such preferences exhibit a form of plasticity.\footnote{Plasticity refers to situations in which an individual's preferences depend not only on the action profile in the underlying game but also on her partner's preference type. Studies of preference evolution allowing for plasticity include \citet{sethi2001preference}, \citet{herold2009evolutionary}, \citet{AlgerLehmann2023WP}, and \citet{Avataneo2026}.} Under this form of plasticity, which we call \textit{parochialism}, agents derive utility exclusively from interactions with others of their own type.\footnote{Parochialism describes a narrow concern for one's own group rather than the broader population. The term has been used in several related senses. \citet{Bernhardetal2006nature} define it as a preference for favoring members of one's own group, whereas \citet{ChoiBowles2007Science} define it as hostility toward other groups. Relatedly, the network literature commonly assumes \emph{homophily}, a direct preference for associating with similar others; see \citet{Jackson2014} for a survey.} Parochialism can therefore be understood as selfishness at the group level. Conditional on interacting with their own kind, these agents then seek to maximize joint material payoff. Types in this class can prevail in evolution for two reasons. First, parochialism fosters positive assortative matching, since in any stable outcome, agents with such preferences do not benefit from partnerships with agents of other types. Second, efficient play within the group guarantees them an average fitness at least as high as that of any competing type. This class therefore reveals a distinct route to evolutionary success---one based on selective association and efficient cooperation within the group.

Next, we turn to the case of incomplete information, under which individuals' preference types are their private information. Following the recent literature on stable matching with incomplete information \citep[see, e.g.,][]{liuetal2014ecma, liu2020AER, Chenhu2020TE, Chenhu2021AEJMicro, Liu2025WP, Wang2022,CatoniniWang2026}, we develop a notion of \textit{incomplete-information stability} suited to our setting. Agents match on the basis of publicly observable \emph{labels} and share correct beliefs about the type distribution associated with each label, but do not directly observe one another's types. As under complete information, internal stability requires self-enforcing play within existing matches, while external stability rules out mutually profitable rematching opportunities. The main conceptual challenge is to define blocking under incomplete information. Intuitively, a blocking proposal identifies a targeted type on each side and prescribes how both sides will play following the deviation. The targeted agents must find it optimal to follow the proposal and must strictly benefit from doing so. Because types are not directly observable, these conditions must continue to hold even if nontargeted types also join the deviation and behave differently from the targeted types. The resulting solution concept has two useful properties. First, it formally incorporates complete-information stability as a special case, thereby guaranteeing existence. Second, information is \emph{endogenous} in a stable outcome, because the informational content conveyed by labels must itself be consistent with stability.

We show that morally constrained selfish types can be at an evolutionary disadvantage in a broad class of material games. Under complete information, agents of such a type can secure a high material payoff: if their current payoff is too low, they can identify one another, rematch, and coordinate on a fair and efficient equilibrium. Under incomplete information, however, the same deviation may no longer be attractive. Because a prospective partner's type is unobservable, a deviating agent must account for the possibility that nontargeted types also participate and respond adversely to the proposed play. The deviation may therefore fail to constitute an incomplete-information blocking pair in an unfavorable outcome, preventing morally constrained selfish agents from guaranteeing an average material payoff at least as high as that of a competing type.

In contrast, parochial efficient types continue to prevail in evolution even under incomplete information. Indeed, agents with such preferences do not face the same obstacle confronting morally constrained selfish types. Because they derive positive utility only from interacting with their own type and none from interacting with competing types, the behavior of nontargeted participants cannot eliminate their incentive to leave an unfavorable match and rematch with one another. We show that every incomplete-information stable outcome exhibits perfect assortative matching by label, and that within-label play is efficient for every label that may contain parochial efficient agents.\footnote{Although our analysis characterizes stable outcomes rather than the adjustment process leading to them, the results admit a heuristic interpretation of partial unraveling. Starting from coarse labels, parochial efficient agents can form a blocking pair whenever the matching is not assortative or their play is inefficient. If participation in and behavior following deviations are publicly observed and informative, rematching can promote assortative matching directly and, at the same time, generate information about agents' underlying types that can be incorporated into finer labels.} Together, these properties ensure that parochial efficient agents earn an average material payoff at least as high as that of any competing type, regardless of its population share.


The two papers most closely related to ours are \citet{Dekeletal2007RESTUD} and \citet{AlgerAndWeibull2013ECMA}. \citet{Dekeletal2007RESTUD} study preference evolution under uniform random matching and an exogenous degree of preference observability. They find that observability shapes which behaviors can survive evolutionary selection. When preference types are perfectly observable, only efficient outcomes can survive; when they are completely unobservable, by contrast, self-interested behavior becomes evolutionarily robust.  They also consider intermediate degrees of observability and show that the necessity of efficient and selfish behaviors continues to hold in neighborhoods of the two polar cases. Our results under complete information similarly emphasize the force toward efficient play. Our main departure, however, arises under incomplete information. With stable partner choice, the information conveyed by observable labels is itself endogenous. We show that parochial efficient types can overcome the informational friction involved in forming blocking pairs, sustain efficient play, and thus prevail in evolution even when types are unobservable.

\citet{AlgerAndWeibull2013ECMA} also study preference evolution under incomplete information but impose an exogenous matching process with a given degree of assortativity. Their central result establishes a sharp link between matching and morality: \emph{homo moralis}, whose utility is a convex combination of own material payoff and a Kantian moral objective, is evolutionarily successful when the weight on the Kantian term equals the degree of assortativity. Our contribution is to derive assortativity endogenously from stable partner choice. Our results reinforce their broader insight that evolution can favor preferences combining self-interest with a moral concern, but the resulting preferences differ in two crucial respects. First, the moral objective is defined differently. Whereas the Kantian criterion in \citet{AlgerAndWeibull2013ECMA} evaluates an action by the payoff it would yield if the partner chose the same action, moral play in our model maximizes joint material payoff. It may therefore require the partners to take different roles and actions, which is made possible by coordination within a match. Second, our types combine self-interest and the moral concern in a lexicographic manner: the lower-priority concern matters only when the higher-priority concern is satisfied.\footnote{Any convex combination of selfishness and material efficiency does not prevail in our model. For every such preference type, there exists a material game in which it is evolutionarily disadvantaged.}


\subsection{Related Literature}


\citet{Frank1987} studies preference evolution with endogenous matching under incomplete information. His model imposes an exogenous information structure that induces positive assortative matching, after which matched agents interact only once. Frank shows that commitment power and the ability to identify committed partners are key drivers of evolutionary success. Our results operate through a related mechanism. Preferences that induce efficient play serve as a form of conditional commitment, while parochialism combined with stable partner choice allows type identification and assortative matching to arise endogenously. Subsequent studies by \citet{Nowak2000Fairness}, \citet{Akdeniz2021}, and \citet{Akdeniz2023} build on this insight, showing how evolution can favor commitment to behavior that would otherwise be irrational.
 
In a random-matching environment under incomplete information, \citet{GuthKliemt1998RS} demonstrate that conditional cooperators cannot survive when preference types are unobservable, because they cannot tailor their behavior to their opponents' types. Subsequent contributions include \citet{ElyYilankaya2001JET}, \citet{OkVega-Redondo2001JET}, and \citet{Dekeletal2007RESTUD}. More recently, \citet{AvataneoNormanPersico2025} show that the five principles of Moral Foundations Theory generically span the set of evolutionarily stable preferences under random matching and observable types. Both \citet{herold2009evolutionary} and \citet{Avataneo2026} allow plastic preferences which depend on the opponent's type, as we do in this paper. \citet{herold2009evolutionary} show that discriminating types are evolutionarily stable under (almost) complete information, while \citet{Avataneo2026} finds that universalism is favored when repeated interaction with the same partner is unlikely and particularism when it is likely. We also recognize the importance of plasticity but show that it plays a different role under endogenous matching: it shapes not only how agents behave toward different types, but also whom they choose to interact with and, consequently, the resulting matching patterns.

\citet{AlgerAndWeibull2013ECMA} study preference evolution under incomplete information and exogenous matching with a given degree of assortativity. \citet{newton2017IJGT} extends \citet{AlgerAndWeibull2013ECMA} by allowing the degree of matching assortativity to evolve. He demonstrates that Kantian preferences can prevail when coupled with parochialism defined through the matching rule. \citet{Wu2019IJGT} studies observable labels that are exogenously correlated with otherwise unobservable preference types and assumes homophilic matching on those labels. In contrast, our model embeds parochialism directly in preferences and endogenizes, through stable partner choice, both the information conveyed by labels and the resulting degree of assortativity. Our approach is also related to models of endogenous interaction structures and dynamic partner choice \citep[see][among others]{JacksonWatts2002,Fujiwara-GrevandOkuno-Fujiwara2009,Izquierdoetal2010,Izquierdoetal2014,Izquierdo2021,Graser2024}.


\section{Population, Material Game, and Preference Types} \label{sec: model}
Consider a continuum population of unit mass. Agents are matched in pairs and interact through a symmetric game $\G$. The game has a finite common action set $A$. If an agent plays action $a\in A$ while her partner plays action $a'\in A$, she receives a \textbf{material payoff} (or \textbf{fitness}) $\pi(a,a')$, where $\pi:A^2\to\mathbb{R}$. Because $A^2$ is finite and every agent participates in exactly one match, adding the same constant to all material payoffs does not affect fitness comparisons. We therefore normalize $\pi(a,a')>0$ for all $(a,a')\in A^2$.\footnote{We do not allow agents to opt out of playing the material game. Nevertheless, our analysis does extend to a setting in which agents can remain unmatched, provided that the outside option yields a material payoff strictly below the lowest payoff attainable from any match---for example, zero under our normalization.}

Write $\Theta$ for the set of \textbf{preference types}. Formally, a preference environment is a pair $(\Theta,u)$, where
\[
u: A^2 \times  \Theta\times\Theta\rightarrow\mathbb{R}.
\]
For $\theta,t\in\Theta$, write $u_\theta(a,a',t)= u(a,a',\theta,t)$ for the utility of a type-$\theta$ agent who plays $a$ against a type-$t$ partner who plays $a'$. We interpret $\Theta$ as the set of preference types that may arise through mutation and assume that it includes all types constructed in the following analysis.\footnote{This reduced-form specification follows \citet{herold2009evolutionary} \citep[see also][]{Avataneo2026}. \citet{gul2016} provide a foundation for such interdependent preferences by constructing a canonical type space from characteristics and hierarchies of preference responses.} This specification extends the standard outcome-based formulation in the literature on preference evolution by allowing utility to depend on the matched partner's preference type. We call type $\theta$ \textbf{plastic} if its utility depends on its partner's type; that is, if $u_\theta(a,a',t)\neq u_\theta(a,a',t')$ for some $t,t'\in\Theta$ and $(a,a')\in A^2$.

We make two additional remarks. First, we focus on material games with a finite common action set $A$, as in \citet{Dekeletal2007RESTUD}. Our analysis extends to games with a general topological strategy space under standard regularity conditions, but the central insights of the paper remain unchanged. Second, we impose no restriction linking the material payoff function $\pi$ to the utility function $u_\theta$. Thus, agents may maximize objectives other than their own material fitness. The following example illustrates the flexibility allowed by this formulation.

\begin{example}\label{ex: utility-examples}
The model accommodates a wide range of preferences.
\begin{itemize}
    \item[(i)] A \textbf{selfish} type cares only about her own material payoff: $u_\theta(a,a',t)=\pi(a,a')$.
    \item[(ii)] An \textbf{efficient} type cares about the total material payoff generated within the matched pair: $u_\theta(a,a',t)=\pi(a,a')+\pi(a',a)$.
    \item[(iii)] A \textbf{behavioral} type may have an objective unrelated to material payoffs. For example, for some fixed action $\bar a\in A$, $u_\theta(a,a',t)=\mathbf{1}_{\{a=\bar a\}}$.
    \item[(iv)] A \textbf{plastic} type conditions its concern for the partner's payoff on the partner's type. For example, $u_\theta(a,a',t)=\pi(a,a')+\pi(a',a)\cdot\mathbf{1}_{\{t=\theta\}}$.\qedhere
\end{itemize}
\end{example}

When two agents interact, they may coordinate their behavior through an \textbf{agreement} $\alpha\in\Delta(A^2)$, defined as a probability distribution over action pairs. Let $\mathcal A=\Delta(A^2)$ denote the set of all possible agreements. We impose no restrictions on $\alpha$, so agreements may involve arbitrary correlation between agents' actions. This formulation is deliberately general. An agreement may be deterministic, placing probability one on a single action pair $(a,a')$. It may also specify an independent joint distribution, corresponding to the agents' use of mixed strategies. More generally, agents may coordinate their actions through a randomizing device that generates public or private action recommendations.

Randomizing devices are common in real-world interactions and provide a natural way for agents to coordinate behavior. This is particularly relevant in an evolutionary setting, where agents interact over a long horizon and may learn to utilize such devices. Consider the familiar Battle of the Sexes, in which agents benefit from coordination but disagree about which activity to coordinate on.\footnote{Formally, the Battle of the Sexes can be represented as a symmetric two-player game with a common action set by defining actions in player-relative terms: each player chooses either her own preferred activity or her partner's preferred activity.} A public coin flip can eliminate miscoordination while treating the agents symmetrically ex ante. It can also be viewed as a reduced-form representation of a long-run arrangement such as turn-taking.

For our main analysis, we consider a population with two distinct preference types $\theta,\tau\in\Theta$.\footnote{In Section \ref{sec: polymorphism}, we discuss how our results extend to the more general case of polymorphic populations.} A proportion $1-\varepsilon$ of the agents are of type $\theta$, and the remaining proportion $\varepsilon$ are of type $\tau$, where $\varepsilon\in(0,1)$. We refer to the tuple $(\theta,\tau,\varepsilon)$ as a \textbf{population state}.


\section{Preference Evolution with Complete Information} \label{sec: preference evolution complete info}

In this section, we assume that each agent observes the preference types of all other agents. Hence, when two agents are matched, they play $\G$ with complete information. 

Fix a population state $(\theta, \tau, \varepsilon)$. For each type $t\in \{\theta, \tau\}$, we let $\mu_{t} \in \Delta(\{\theta, \tau\})$ be a probability distribution over types in the population that describes how type-$t$ agents are matched. A \textbf{matching profile} is a vector $\mu=(\mu_{\theta},\mu_{\tau})$ that satisfies the following consistency condition:
\[
(1-\varepsilon)\mu_{\theta}[\tau]=\varepsilon \mu_{\tau}[\theta].
\]
This condition requires that the total mass of type-$\theta$ agents matched with type-$\tau$ agents is equal to that of type-$\tau$ agents matched with type-$\theta$ agents.\footnote{Although we take a distributional approach in defining the matching profile, there is no randomness in how agents are matched. Given that the population consists of finitely many types, we can explicitly describe the matching pattern through a deterministic mapping that generates $\mu$.}

For any $t,t'\in\{\theta,\tau\}$, let $\gamma_{t,t'} \in \Delta(\A)$ describe the conditional distribution of agreements reached across matches between type-$t$ and type-$t'$ agents. Let $\rho: \A\rightarrow \A$ be a mapping that reverses the roles of the two agents; that is, for any $\alpha\in\A$,
\[
\rho(\alpha)[a,a']=\alpha[a',a] \ \ \text{for all $(a,a')\in A^2$}.\]
An \textbf{agreement profile} $\gamma=(\gamma_{t,t'})$ is a vector of conditional distributions of agreements that satisfies the following exchangeability condition: $\gamma_{t,t'}[B]=\gamma_{t',t}[\rho(B)]$ for any measurable set $B\subseteq \A$. 

We call the combination of a matching profile and an agreement profile $(\mu, \gamma)$ an \textbf{outcome}. 


\subsection{Stable Outcome}\label{subsec: stable outcome with ci}
Given a population state $(\theta, \tau, \varepsilon)$, our next goal is to identify the outcomes $(\mu, \gamma)$ that can be deemed \textit{stable}. The requirement of stability has two layers. First, holding the matching profile $\mu$ fixed, agents do not want to change their behavior as specified by $\gamma$. In other words, every agreement should induce equilibrium play. Second, given the utilities agents derive in an outcome, there should not exist agents who want to form a pairwise deviation and mutually benefit from rematching. To formalize this idea, we extend the stability notion of \citet{Garrido-LucerolLaraki2021WP} to a continuous population and from Nash equilibrium to correlated equilibrium; see also \citet{JacksonWatts2010}. 


We now formally define these two layers of stability. For each agreement $\alpha\in\A$ between $t$ and $t'$ and a function $g:A\rightarrow A$, let
\[
\mathbb{E}_\alpha[u_t(g(a),a',t')]=\sum_{(a,a')\in A^2} u_t(g(a),a',t')\alpha[a,a']
\]
denote the expected utility type-$t$ agent obtains by playing $g(a)$ whenever the agreement recommends action $a$. Let $f:A \rightarrow A$ be the strategy of \emph{following} the recommendation so that $f(a)=a$ for all $a\in A$. For $t,t'\in\{\theta,\tau\}$, let $\CE_{t,t'}$ denote the set of correlated equilibria between type-$t$ and type-$t'$ agents. That is, $\alpha\in\CE_{t,t'}$ if and only if
\[
f\in\argmax_g \, \mathbb{E}_{\alpha} [u_t(g(a),a',t')] \quad\text{and}\quad f\in\argmax_g \, \mathbb{E}_{\rho(\alpha)} [u_{t'}(g(a),a',t)].
\]

As discussed in Section \ref{sec: model}, an agreement can be interpreted as a correlation device that draws an action pair and privately recommends one action to each agent. A correlated equilibrium requires the agreement to be self-enforcing: conditional on her own recommendation, each agent finds it optimal to follow it when her partner does the same \citep{Aumann1974,Aumann1987}. This solution concept is natural in our setting. Matched agents can coordinate before play but cannot commit to their actions. Correlated equilibrium captures precisely this combination: it permits coordination while requiring its recommendations to be voluntarily followed.\footnote{The direct-recommendation formulation is without loss of generality. More generally, agents could agree ex ante on an arbitrary payoff-irrelevant signal structure and play a Bayes-Nash equilibrium conditional on their signals. The resulting distribution of action pairs is a correlated equilibrium \citep{Aumann1987}.}

An agreement profile is a correlated equilibrium profile if almost every matched pair plays a correlated equilibrium.

\begin{definition}\label{def: correlated equilibrium profile}
An agreement profile $\gamma$ is a \textbf{correlated equilibrium profile} with respect to $\mu$ if for every $t,t'\in\{\theta,\tau\}$ such that $\mu_t[t']>0$, $\gamma_{t,t'}[\CE_{t,t'}]=1$.
\end{definition}

A block specifies two agents who are present in matches with positive mass, together with an agreement for their rematch. The block is viable under complete information if the proposed agreement is self-enforcing for the two deviating types and makes both of them strictly better off compared to the status quo. For $t\in\{\theta,\tau\}$, call a pair $(\bar{t},\bar{\alpha})\in\{\theta,\tau\}\times\A$ a \textbf{current position} of type $t$ under $(\mu,\gamma)$ if $\mu_t[\bar{t}]>0$ and $\bar{\alpha}\in\supp(\gamma_{t,\bar{t}})$.

\begin{definition}\label{def: complete information blocking}
    Fix an outcome $(\mu,\gamma)$. A \textbf{complete-information blocking pair} exists if there are types $t,t'\in\{\theta,\tau\}$; current positions $(\bar t,\bar\alpha)$ of type $t$ and $(\bar{t}',\bar{\alpha}')$ of type $t'$; and an agreement $\hat\alpha\in\A$ such that
    \begin{itemize}
    \setlength{\itemsep}{0pt}
    \item[(i)] $f\in\argmax_{g}\,\mathbb{E}_{\hat{\alpha}}[u_t(g(a),a',t')]$ and $f\in\argmax_{g}\,\mathbb{E}_{\rho(\hat{\alpha})}[u_{t'}(g(a),a',t)]$;
    \item[(ii)] $\mathbb{E}_{\hat{\alpha}}[u_t(a,a',t')] > \mathbb{E}_{\bar{\alpha}}[u_t(a,a',\bar{t})]$ and $\mathbb{E}_{\rho(\hat{\alpha})}[u_{t'}(a,a',t)] > \mathbb{E}_{\bar{\alpha}'}[u_{t'}(a,a',\bar{t}')]$.
    \end{itemize}
\end{definition}

This notion of blocking for aggregate matching is analogous to the one in \citet{Echeniqueetal2013ECMA}, which is a natural generalization of the blocking concept proposed by \citet{GaleShapley1962AMM} to continuous populations. For a continuous population, deviating agents in a blocking pair must have positive mass; otherwise, they would be unable to locate one another. Moreover, the proposed agreement must satisfy two requirements. First, it must be self-enforcing: the deviating agents coordinate on a correlated equilibrium, so following the action recommendation is optimal for each side. Second, it must be profitable: both agents strictly prefer the proposed rematch to their current positions.\footnote{Our stability notion requires both parties to have strict incentives to rematch. This requirement is standard in the literature on matching games without transfers, as it is robust to small payoff perturbations or infinitesimal rematching costs. A weaker notion of blocking that imposes a weak incentive on one side can sharpen the set of stable outcomes but is known to generate unappealing non-existence issues. We use the strict-gain formulation as a conservative and robust test for whether a proposed rematch should count as a blocking pair.}




For an outcome to be stable, almost every matched pair should play a correlated equilibrium, and no blocking pair should be able to upset the outcome. This leads to the following definition.

\begin{definition}\label{def: stable outcome}
An outcome $(\mu, \gamma)$ is \textbf{complete-information stable} if it satisfies:
\begin{itemize}
\item[(i)] (Internal stability) $\gamma$ is a correlated equilibrium profile with respect to $\mu$;
\item[(ii)] (External stability) There is no complete-information blocking pair.
\end{itemize}
\end{definition}

Complete-information stability describes outcomes that, once reached, leave no room for further strategic or coalitional adjustment. In the spirit of pairwise stability in matching games, agents evaluate deviations myopically: they compare the one-shot payoff from a proposed rematch with their current payoff, without considering how the deviation may affect future opportunities.



Since our model has no predetermined sides, the matching problem resembles a ``roommate problem,'' where stable matchings need not exist in finite economies \citep{GaleShapley1962AMM}. In our continuum environment, however, stable outcomes always exist---a result we prove in Proposition \ref{prop: existence} for the more general case of polymorphic populations. The next example illustrates Definition \ref{def: stable outcome}.

\begin{example}\label{eg: complete-info stability}
Consider the following underlying material game and a population state $(\theta, \tau, \varepsilon)$.
\begin{center}
    \begin{tabular}{c|c|c|}
        \multicolumn{1}{c}{} & \multicolumn{1}{c}{$x$}  & \multicolumn{1}{c}{$y$} \\\cline{2-3}
     $x$ & $\  1,1 \ $ & $\ 4,2 \ $ \\\cline{2-3}
       $y$ & $\  2,4 \ $ & $\ 1,1 \ $ \\\cline{2-3}
    \end{tabular}
\end{center}
    


Assume that type $\theta$ is selfish, i.e.,  $u_\theta(a, a', t)=\pi(a, a')$, whereas the preference type $\tau$ is behavioral, with utility function $u_\tau(a, a', t)=\mathbf{1}_{\{a=x\}}$. Since action $x$ is dominant for type-$\tau$ agents, the unique correlated equilibrium $\hat{\alpha}$ between two type-$\tau$ agents satisfies $\hat{\alpha}[x,x]=1$.

We now argue that any complete-information stable outcome $(\mu,\gamma)$ must satisfy $\mu_{\theta}[\theta]=\mu_{\tau}[\tau]=1$; $\gamma_{\theta,\theta}[\tilde{\alpha}]=1$, where $\tilde{\alpha}[x,y]=\tilde{\alpha}[y,x]=1/2$; $\gamma_{\tau,\tau}[\hat{\alpha}]=1$.\footnote{The complete-information stable outcome is unique in a generic sense: $\gamma_{\theta,\tau}$ may be specified arbitrarily for a measure-zero set of cross-type matches and has no bearing on complete-information stability.} That is, matching is perfectly assortative. Every matched pair of type-$\theta$ agents plays $(x,y)$ and $(y,x)$ with equal probability, which can be implemented as a correlated equilibrium with public randomization. Each matched pair of type-$\tau$ agents plays $(x,x)$.

To see this, suppose the contrary. There are two cases to consider:
\begin{itemize}
\item Suppose $\mu_{\theta}[\tau]>0$. Internal stability then implies that the unique agreement $\alpha \in \supp(\gamma_{\theta,\tau})$ satisfies $\alpha[y,x]=1$, so type-$\theta$ agents in cross-type matches obtain utility $2$. But any two such type-$\theta$ agents can block by rematching with each other and playing the correlated equilibrium $\tilde{\alpha}$, which gives each of them expected utility $3$. This contradicts external stability.

\item Suppose $\mu_{\theta}[\theta]=\mu_{\tau}[\tau]=1$, but $\supp (\gamma_{\theta, \theta}) \ne \{\tilde{\alpha}\}$. Then a positive mass of type-$\theta$ agents obtain expected utility strictly below $3$: in any correlated equilibrium between two type-$\theta$ agents, the \emph{lower} of the two agents' expected utilities is at most $3$, and this bound is attained only at $\tilde{\alpha}$. These agents can therefore block by rematching with one another and playing $\tilde{\alpha}$, again violating external stability. \qedhere
\end{itemize}
\end{example}

Example~\ref{eg: complete-info stability} illustrates an \emph{equilibrium-selection} effect of stability. There are many correlated equilibria between two type-$\theta$ agents, but only $\tilde{\alpha}$ can arise in a complete-information stable outcome. External stability rules out the others: if a correlated equilibrium leaves some type-$\theta$ agents below an equal split of the efficient total payoff, those agents can profitably rematch and play $\tilde{\alpha}$.\footnote{This effect is similar in spirit to that analyzed by \citet{JacksonWatts2010}. In both settings, stability restricts the outcome and thereby refines the set of equilibria that can arise.}

\subsection{Evolutionary Dominance}\label{sec: evo dominance}

Following the indirect evolutionary approach, we assume that, at the behavioral level, the population reaches a stable outcome before evolutionary forces appreciably alter the distribution of types. At the preference level, the type distribution evolves according to the material payoffs earned by different types in stable outcomes. Importantly, stable outcomes are determined by agents' subjective preferences, whereas evolutionary selection is governed by material success.

Given a complete-information stable outcome $(\mu,\gamma)$ in population state $(\theta,\tau,\varepsilon)$, the average material payoffs for type-$\theta$ and type-$\tau$ agents are given by
\begin{align*}
G_\theta(\mu,\gamma)&=\sum_{t\in\{\theta,\tau\}}\left[\int_{\A}\mathbb{E}_{\alpha}[\pi(a,a')]\mathop{d \gamma_{\theta,t}[\alpha]}\right] \mu_{\theta}[t],\\
G_\tau(\mu,\gamma)&=\sum_{t\in\{\theta,\tau\}}\left[\int_{\A}\mathbb{E}_{\alpha}[\pi(a,a')]\mathop{d \gamma_{\tau,t}[\alpha]}\right] \mu_{\tau}[t].
\end{align*}

We now define a static notion that captures the evolutionary selection process, called evolutionary dominance, as follows.

\begin{definition}\label{def: evo dominance}
Fix a material game $\G$.
\begin{itemize}
    \item[(i)] A preference type $\theta \in \Theta$ is \textbf{evolutionarily dominant} if for every $\tau \in \Theta$ and every $\varepsilon\in(0,1)$, every complete-information stable outcome $(\mu,\gamma)$ in the population state $(\theta,\tau,\varepsilon)$ satisfies $G_\theta(\mu, \gamma)\geq G_\tau(\mu, \gamma)$. 
    \item[(ii)] A preference type $\theta\in\Theta$ is \textbf{evolutionarily dominated} if there exists another type $\tau\in\Theta$ such that for every $\varepsilon\in(0,1)$, every complete-information stable outcome $(\mu,\gamma)$ in the population state $(\theta,\tau,\varepsilon)$ satisfies $G_\theta(\mu,\gamma)\leq G_\tau(\mu,\gamma)$, with strict inequality for some complete-information stable outcome.
\end{itemize}
\end{definition}

\begin{remark}
Our definition of evolutionary dominance is closely related to the notions of evolutionary stability in \citet{Dekeletal2007RESTUD} and \citet{AlgerAndWeibull2013ECMA}, but it differs in two respects. First, those papers fix an exogenous matching technology and compare average material payoffs across equilibrium play; in contrast, we endogenize matching and compare payoffs across stable outcomes. Second, their notions are defined in a local sense, while ours is global: the payoff comparison should hold for every population share $\varepsilon$ of the competing type $\tau$. Hence, an evolutionarily dominant type $\theta$ must both resist invasion when it is the incumbent type (i.e., $\varepsilon$ is close to $0$) and have the ability to invade when it is the mutant type (i.e., $\varepsilon$ is close to $1$).
\end{remark}

Two natural benchmarks---the selfish and efficient types---capture two forces central to evolutionary success, but neither is robust on its own. Selfish preferences protect agents from exploitation but may sustain inefficient play, whereas efficient preferences promote material efficiency but may expose agents to unfavorable payoff divisions. We defer a formal analysis of these limitations to Section \ref{sec: pure types} and first introduce two new classes of preference types that combine selfishness and efficiency nonlinearly. One subjects selfish behavior to an efficiency constraint; the other makes efficiency concerns conditional on interacting with one's own type. Our main result under complete information is that types in both classes are evolutionarily dominant.

Let $m$ denote the maximum joint material payoff generated by a pure action pair; that is
    \[
        m=\max_{(a, a') \in A^2}\,\pi(a,a')+\pi(a',a).
    \]
Moreover, let $\E\subseteq A^2$ be the set of action pairs that attain this joint payoff:
    \[
        \E=\left\{(a, a') \in A^2: \pi(a,a')+\pi(a',a)=m \right\}.
    \]
An agreement $\alpha$ is \textbf{efficient} if it assigns probability one to $\E$, i.e., $\alpha[\E]=1$; or equivalently, $\mathbb{E}_{{\alpha}}[\pi(a,a')+\pi(a',a)]=m$.\footnote{In previous literature under exogenous matching, since agents of the same type are assumed to play the same action in equilibrium, the consideration of efficiency is restricted to symmetric action profiles. See, for example, \citet{Dekeletal2007RESTUD}.} An agreement that is not efficient is \textbf{inefficient}.

\begin{definition}\label{def: morally constrained selfish type}
A preference type $\theta$ is \textbf{morally constrained selfish} if, up to a positive affine transformation, its utility function can be written as
\begin{equation}\label{eq: morally constrained selfish}
    u_\theta(a,a',t)
    =
    \mathbf{1}_{\{(a,a')\in\E\}}\cdot\pi(a,a').
\end{equation}
\end{definition}

Because positive affine transformations preserve expected-utility comparisons, we henceforth adopt the normalization in (\ref{eq: morally constrained selfish}) without loss of generality. We use the same convention for all preference classes defined below.

Recall that $\E$ is the set of efficient action pairs. A morally constrained selfish type then embodies a hierarchical ordering of objectives: agents with this preference type first seek efficient interaction with their partner and then, among efficient outcomes, choose the action that maximizes their own material payoff. We have the following positive result:

\begin{proposition}\label{prop: morally constrained selfish}
For any material game $\G$, a morally constrained selfish type is evolutionarily dominant.
\end{proposition}

The intuition for Proposition~\ref{prop: morally constrained selfish} is as follows. Let $\theta$ denote a morally constrained selfish type. The moral constraint ensures that there exists an efficient agreement $\alpha\in \CE_{\theta,\theta}$, so two type-$\theta$ agents can sustain efficient play (Lemma \ref{lemma: efficient CE}). The selfish component and external stability then guarantee that almost every type-$\theta$ agent obtains expected material payoff at least $m/2$. For if not, two such agents can form a blocking pair by playing $\alpha/2+\rho(\alpha)/2$: such an agreement delivers $m/2$ and also belongs to $\CE_{\theta,\theta}$ by symmetry and convexity of the set of correlated equilibria. Because no matched pair can generate total material payoff above $m$, no other type can earn an average material payoff above $m/2$ when type $\theta$ earns no less than $m/2$. The morally constrained selfish type therefore performs at least as well as any competing type in every stable outcome.

\begin{definition} \label{def: parochial efficient}
A preference type $\theta$ is \textbf{parochial efficient} if, up to a positive affine transformation, its utility function can be written as
\begin{equation}\label{eq: parochial efficient}
u_\theta(a, a', t)=[\pi(a,a')+\pi(a',a)]\cdot\mathbf{1}_{\{t=\theta\}}.
\end{equation}
\end{definition}

The utility function (\ref{eq: parochial efficient}) of a parochial efficient type also has a hierarchical structure. At the first level, such agents derive utility only from interacting with partners who share their preference type, a form of plasticity we call \textit{parochialism}. Their first-order objective is therefore to sort with their own type. We interpret this parochial concern as selfishness at the group level. Conditional on same-type matching, agents then seek to maximize joint material payoff. 

\citet{newton2017IJGT} also considers parochialism in preference evolution. He defines it at a matching level: parochial agents are assumed to match only with one another. In contrast, we define parochialism at the preference level. Whether parochial agents actually sort with their own type is determined endogenously by stable partner choice.

We show that a parochial efficient type can also prevail in the long run. 

\begin{proposition}\label{prop: parochial efficient}
For any material game $\G$, a parochial efficient type is evolutionarily dominant.
\end{proposition}

The mechanism behind Proposition~\ref{prop: parochial efficient} differs from that behind Proposition~\ref{prop: morally constrained selfish}. The evolutionary dominance of a parochial efficient type rests on two behavioral implications: assortative matching and efficient play. The parochial component gives agents an incentive to match with their own type, while the efficiency component leads them to maximize joint material payoff within same-type matches. Assortative matching protects these agents from exploitation by other types. Thus, the vulnerability of a purely efficient type (see Example~\ref{example: efficient failure} below) is neutralized: when efficient play requires one agent to accept a lower material payoff, the corresponding gain accrues to another agent with the same preferences and therefore benefits the same evolutionary lineage.


While the preceding results identify nonlinear combinations of efficiency and selfishness that are sufficient for evolutionary dominance, they do not exhaust the preference types favored by evolutionary pressure. Nevertheless, the following result provides a necessary condition for a preference type to prevail.


\begin{proposition}\label{prop: necessary with ci}
Fix a material game $\G$ and a preference type $\theta$. If there exists an agreement
\[\hat{\alpha}\in\argmax_{\alpha\in \CE_{\theta,\theta}}\,\mathbb{E}_\alpha[u_{\theta}(a,a',\theta)+u_{\theta}(a',a,\theta)]\]
that is inefficient, then $\theta$ is evolutionarily dominated.
\end{proposition}

To understand this result, first observe that any correlated equilibrium has a fair counterpart with the same joint expected utility: assigning its two roles symmetrically equalizes the agents' expected utilities while preserving their incentives to follow its recommendations. Moreover, recall that stable partner choice has an equilibrium-selection effect. In same-type matches, only correlated equilibria that are fair and attain the highest joint expected utility can be sustained. Otherwise, two agents in disadvantageous positions can form a blocking pair. This selection effect, however, operates according to agents' preferences rather than material payoffs. If an inefficient correlated equilibrium attains the highest joint expected utility, its fair version remains inefficient. Materially inefficient play can therefore persist in same-type matches. Consequently, such a preference type is evolutionarily dominated by a competing type that can force assortative matching and efficient play.

An implication of Proposition \ref{prop: necessary with ci} is that, if a population consists of a single type that is evolutionarily dominant, then almost every pairwise interaction must be efficient in a stable outcome (see a generalization of this insight in Proposition \ref{prop: necessary condition for polymorphism}(ii)). This conclusion is consistent with that of \citet{Dekeletal2007RESTUD}, obtained under complete information in a random-matching environment.

\subsection{Pure Selfishness and Efficiency}\label{sec: pure types}

We now return to two familiar benchmarks: the selfish and efficient types. Up to positive affine transformations, their utility functions are, respectively, $u_\theta(a,a',t)=\pi(a,a')$ and $u_\theta(a,a',t)=\pi(a,a')+\pi(a',a)$. These types isolate two key forces emphasized in the literature on preference evolution: resistance to exploitation and the ability to sustain efficient play. The following proposition shows, however, that neither force alone guarantees evolutionary dominance without additional restrictions on the material game.

We call an agreement a correlated equilibrium of the material game if it is a correlated equilibrium when both agents’ utilities coincide with their material payoffs.

\begin{proposition} \label{prop: negative result}
Either a selfish type or an efficient type may be evolutionarily dominated for some material game $\G$.
\begin{itemize}
    \item[(i)] If there exists an efficient correlated equilibrium of the material game, then a selfish type is evolutionarily dominant; otherwise, a selfish type is evolutionarily dominated.
    \item[(ii)] If $\pi(a,a')=\pi(a',a)$ for every $(a,a')\in\E$, then an efficient type is evolutionarily dominant; otherwise, an efficient type is evolutionarily dominated.
\end{itemize}
\end{proposition}

For a selfish type, the evolutionary weakness lies in its failure to promote material efficiency: selfish agents may become trapped in an inefficient correlated equilibrium. The following example illustrates this possibility.

\begin{example}\label{example: selfish failure}
Suppose that the underlying material game is given as follows:
\begin{center}
    \begin{tabular}{c|c|c|}
        \multicolumn{1}{c}{} & \multicolumn{1}{c}{$x$}  & \multicolumn{1}{c}{$y$} \\\cline{2-3}
     $x$ & $\  1,1 \ $ & $\ 2,8 \ $ \\\cline{2-3}
       $y$ & $\  8,2 \ $ & $\ 3,3 \ $ \\\cline{2-3}
    \end{tabular}    
\end{center}


Let $\theta$ denote a selfish type and $\tau$ a parochial efficient type. Fix any $\varepsilon\in(0,1)$. By Proposition \ref{prop: parochial efficient}, $G_\tau(\mu,\gamma)\geq G_\theta(\mu,\gamma)$ in every complete-information stable outcome. It therefore remains to construct a stable outcome in which the inequality is strict.

Consider the outcome $(\mu,\gamma)$ with perfectly assortative matching, $\mu_\theta[\theta]=\mu_\tau[\tau]=1$.
In type-$\theta$ matches, let $\gamma_{\theta,\theta}[\hat{\alpha}]=1$, where $\hat{\alpha}[y,y]=1$. In type-$\tau$ matches, let $\gamma_{\tau,\tau}[\tilde{\alpha}]=1$, where $\tilde{\alpha}[x,y]=\tilde{\alpha}[y,x]=1/2$.

Internal stability follows because $y$ is strictly dominant for selfish agents, while following $\tilde{\alpha}$ maximizes the joint material payoff of parochial efficient agents matched with their own type. For external stability, every type-$\tau$ agent obtains utility $10$, the highest feasible utility, so no blocking pair can involve type $\tau$. Moreover, $\hat{\alpha}$ is the unique correlated equilibrium between two type-$\theta$ agents and gives each of them the same material payoff as in the status quo. Hence, no two type-$\theta$ agents can form a blocking pair. The outcome is therefore complete-information stable.

In this outcome,
\[
G_\tau(\mu,\gamma)=5>3=G_\theta(\mu,\gamma).
\]
Because $\varepsilon$ was arbitrary, the selfish type $\theta$ is evolutionarily dominated by the parochial efficient type $\tau$.
\end{example}

An efficient type faces the opposite problem: its benevolent concern for joint material payoff leaves it indifferent to how that payoff is divided, which makes it vulnerable to exploitation. To see this, consider the following example.

\begin{example}\label{example: efficient failure}
Suppose that the material game is the same as in Example \ref{example: selfish failure}. Let $\theta$ denote an efficient type and $\tau$ a morally constrained selfish type. By Proposition \ref{prop: morally constrained selfish}, $G_\tau(\mu,\gamma)\geq G_\theta(\mu,\gamma)$ in every complete-information stable outcome. It therefore remains to construct a stable outcome in which the inequality is strict.

Fix any $\varepsilon\in(0,1)$ and choose $c\in(0,\min\{\varepsilon,1-\varepsilon\})$. Consider a matching profile satisfying $(1-\varepsilon)\mu_\theta[\tau]=\varepsilon\mu_\tau[\theta]=c$, with the remaining probability assigned to same-type matches. In cross-type matches, let $\gamma_{\theta,\tau}[\hat{\alpha}]=1$, where $\hat{\alpha}[x,y]=1$. In same-type matches, let $\gamma_{\theta,\theta}[\tilde{\alpha}]=\gamma_{\tau,\tau}[\tilde{\alpha}]=1$, where $\tilde{\alpha}[x,y]=\tilde{\alpha}[y,x]=1/2$.

Internal stability follows because every prescribed action pair is efficient. Type-$\theta$ agents therefore obtain their highest feasible utility, while a type-$\tau$ agent can deviate only by making the action pair inefficient, which gives her utility zero.

For external stability, every type-$\theta$ agent already obtains the highest feasible utility, so no blocking pair can involve type $\theta$. Type-$\tau$ agents obtain expected utility $5$ in same-type matches and utility $8$ in cross-type matches. Because no agreement between two type-$\tau$ agents can generate joint expected utility greater than $10$, no two such agents can both strictly benefit from rematching. The outcome is therefore complete-information stable.

Finally,
\[
G_\tau(\mu,\gamma) = 5\left(1-\frac{c}{\varepsilon}\right)+8\frac{c}{\varepsilon}>5>5\left(1-\frac{c}{1-\varepsilon}\right)+2\frac{c}{1-\varepsilon}=G_\theta(\mu,\gamma).
\]
Because $\varepsilon$ was arbitrary, the efficient type $\theta$ is evolutionarily dominated by the morally constrained selfish type $\tau$.
\end{example}

\begin{remark}
    One may wonder whether any convex combination of selfish and efficient preferences can prevail across material games. The answer is no. Indeed, any nontrivial convex combination yields a preference type with utility function $u_\theta(a,a',t)=\pi(a,a')+\beta\pi(a',a)$ for some $\beta\in(0,1)$. For every such $\beta$, one can construct a material game in which the corresponding type satisfies the condition in Proposition \ref{prop: necessary with ci} and is therefore evolutionarily dominated. Thus, preferences of this form face the same problem as selfish types do and may fail to sustain efficient play in equilibrium.
\end{remark}

\section{Preference Evolution with Incomplete Information} \label{sec: preference evolution incomplete}
In this section, we turn to incomplete information. Suppose that each agent knows her own preference type but may not observe the preference types of other agents in the population.\footnote{An agent may care about certain characteristics of a potential partner, which, for simplicity, we assume uniquely determine the partner's preferences. When these characteristics are readily observable (e.g., physical appearance), a model with complete information suffices. When they are not directly observable (e.g., empathy or a sense of responsibility), however, incomplete information must be taken into account. A richer formulation could treat characteristics and preference types as separate primitives \citep[see, for example, ][]{gul2016}, but we do not pursue that extension here.} For a fixed population state $(\theta,\tau,\varepsilon)$, we therefore generalize the notion of a matching profile to accommodate incomplete information.

A \textbf{matching profile} (with incomplete information) is a tuple $(\Lambda, p, q,\mu)$. The first component $\Lambda$ is a finite set of \textbf{labels} that are publicly observable.\footnote{Labels are purely informational: they convey information about preference types but do not directly enter agents' utilities.} There is a probability distribution with full support $p\in\Delta(\Lambda)$ over the set of labels. For each $\lambda\in\Lambda$, let $q_\lambda\in\Delta(\{\theta,\tau\})$ denote the distribution of preference types conditional on label $\lambda$. We assume $q_\lambda \neq q_{\lambda'}$ whenever $\lambda \neq \lambda'$, reflecting the fact that different labels convey distinct information, and write $q=(q_\lambda)_{\lambda\in\Lambda}$. The pair $(p,q)$ should satisfy the following marginal condition:
\[
\sum_{\lambda\in\Lambda}p[\lambda]q_\lambda[\theta]=1-\varepsilon.
\]
In words, the masses of type-$\theta$ agents with different labels should sum up to their total mass in the population. For brevity, we shall refer to a type-$\theta$ agent with label $\lambda$ as a type-$\theta_\lambda$ agent. As in the case of complete information, for each $\lambda\in\Lambda$, let $\mu_\lambda\in\Delta(\Lambda)$ be a probability distribution over labels describing how label-$\lambda$ agents are matched. The last component of a matching profile is then the vector $\mu=(\mu_\lambda)_{\lambda\in\Lambda}$, which satisfies the consistency condition:
\[
    p[\lambda] \mu_{\lambda}[\lambda']=p[\lambda'] \mu_{\lambda'}[\lambda] \ \ \text{for all $\lambda,\lambda'\in\Lambda$.} 
\]

Given a matching profile $(\Lambda, p, q, \mu)$, for each $\lambda,\lambda'\in \Lambda$, we let $\gamma_{\lambda,\lambda'}\in \Delta(\A)$ describe the conditional distribution of agreements reached across matches between label-$\lambda$ and label-$\lambda'$ agents. An \textbf{agreement profile} (with incomplete information) $\gamma=(\gamma_{\lambda,\lambda'})$ is a collection of such conditional distributions satisfying the same exchangeability condition as in the case of complete information. We assume that agreements in the current outcome are observable and that any information they reveal is already encoded in the labels. Formally, conditional on the matched agents' labels, the observed agreement conveys no additional information about their preference types. Thus, for any $\lambda,\lambda'\in\Lambda$, observing any agreement in the support of $\gamma_{\lambda,\lambda'}$ leaves the posterior type distribution of the label-$\lambda$ agent equal to $q_\lambda$ and, analogously, that of the label-$\lambda'$ agent equal to $q_{\lambda'}$.\footnote{If observing an agreement induced a posterior over a label-$\lambda$ agent's type different from $q_\lambda$, the label system could be refined so that the resulting posterior corresponded to a separate label. The only restriction imposed by this representation is therefore the finiteness of $\Lambda$, which limits the granularity of the information encoded by labels. A fully general formulation could allow a continuum of labels, but doing so would add technical complexity without offering additional economic insights.}

As before, the combination of a matching profile and an agreement profile $(\Lambda, p, q, \mu, \gamma)$ is called an \textbf{outcome} (with incomplete information). As is standard in the literature on matching under incomplete information, we assume that all agents in the population correctly understand the status quo outcome.

\begin{remark}\label{rmk: endogenous info}
    Because the informational component $(\Lambda,p,q)$ is itself part of the outcome, information is \emph{endogenous} under the notion of stability defined below \citep[see also, e.g.,][]{liuetal2014ecma, liu2020AER, Chenhu2020TE, Chenhu2021AEJMicro, Liu2025WP, Wang2022}. In this paper, we impose no exogenous informational restrictions and consider the set of all outcomes with labels that encode arbitrary information. Such restrictions can nevertheless be incorporated by specifying, for example, a commonly understood signal structure and restricting attention to outcomes consistent with it.\footnote{For illustration, consider a population state $(\theta,\tau,\varepsilon)$ and suppose that information about agents' types is generated before matching according to a signal structure $(\xi_\theta,\xi_\tau)$. For an agent of type $t\in\{\theta,\tau\}$, a publicly observed signal that perfectly reveals her preference type is generated with probability $\xi_t>0$; with the remaining probability $1-\xi_t$, no signal is observed. An outcome $(\Lambda,p,q,\mu,\gamma)$ consistent with this signal structure must satisfy the following additional conditions: there exist $\lambda,\lambda'\in\Lambda$ such that $q_\lambda[\theta]=1$, $q_{\lambda'}[\tau]=1$, $p[\lambda]\geq(1-\varepsilon)\xi_\theta$, and $p[\lambda']\geq\varepsilon\xi_\tau$. Intuitively, these conditions require the outcome to preserve the information disclosed by the initial signal structure while allowing labels to endogenously reveal additional information.} All of our results continue to hold under such exogenous informational restrictions, provided they do not fully disclose all information---that is, as long as some incomplete information remains.
\end{remark}

\subsection{Stable Outcome with Incomplete Information}
For $t\in\{\theta,\tau\}$ and $\lambda\in\Lambda$, we write
\begin{equation*}
\mathbb{E}_{q_\lambda}[u_t(a,a',\cdot)]= u_t(a,a',\theta) q_{\lambda}[\theta] +  u_t(a,a',\tau) q_{\lambda}[\tau],
\end{equation*}
which is the expected utility of a type-$t$ agent from playing $(a, a')$ with a label-$\lambda$ partner. For $\lambda,\lambda'\in\Lambda$, let $\CE_{\lambda,\lambda'}$ denote the set of incomplete-information correlated equilibria between label-$\lambda$ and label-$\lambda'$ agents; formally, $\alpha\in \CE_{\lambda,\lambda'}$ if and only if
\begin{align*}
    f &\in \arg\max_{g}\mathbb{E}_\alpha[\mathbb{E}_{q_{\lambda'}}[u_t(g(a),a',\cdot)]]\ \ \text{for each $t\in \supp(q_\lambda)$}\\ 
    f &\in \arg\max_{g}\mathbb{E}_{\rho(\alpha)}[\mathbb{E}_{q_\lambda}[u_{t'}(g(a),a',\cdot)]]\ \  \text{for each $t'\in \supp(q_{\lambda'})$}.
\end{align*}
In words, when facing a label-$\lambda'$ partner, each label-$\lambda$ agent finds it optimal to follow the agreement, given the belief $q_{\lambda'}$ about the partner's underlying type. The second condition imposes the analogous obedience requirement on every type of the label-$\lambda'$ agent.\footnote{This notion differs slightly from the standard notion of correlated equilibrium under incomplete information \citep[see, e.g.,][]{Forges93,Forges06,Liu15, BergemannMorris16}. In our setting, agents observe and understand the agreement played in their match, and any information conveyed by that observation is already incorporated into their beliefs through the labels. In this respect, our notion of incomplete-information correlated equilibrium resembles the rational expectations equilibrium studied by \citet{Koh2023}.}

\begin{definition}\label{def: incomplete-information correlated equilibrium profile}
With incomplete information, an agreement profile $\gamma$ is a \textbf{correlated equilibrium profile} with respect to $\mu$ if for every $\lambda,\lambda'\in \Lambda$ such that $\mu_\lambda[\lambda']>0$, $\gamma_{\lambda,\lambda'}[\CE_{\lambda,\lambda'}]=1$.
\end{definition}


Because agents observe only the labels of potential partners, rather than their preference types, we require an additional definition to properly formulate pairwise deviations under incomplete information.

\begin{definition}
For a label $\lambda\in\Lambda$, a tuple $(\lambda,D,d)$ is a \textbf{deviation plan} if (i) $D$ is a nonempty subset of $\supp(q_\lambda)$ and (ii) $d:D\rightarrow A^A$.
\end{definition}

In words, $D$ is the set of preference types among label-$\lambda$ agents that participate in a pairwise deviation, while $d$ specifies their behavior. For each $t\in D$, $d(t):A\rightarrow A$ is the strategy used in the deviation: upon receiving recommendation $a$, a type-$t$ agent takes action $d(t)(a)$. As in the complete-information case, we call a pair $(\bar{\lambda},\bar{\alpha})\in\Lambda\times\A$ a \textbf{current position} of a label-$\lambda$ agent under the outcome $(\Lambda,p,q,\mu,\gamma)$ if $\mu_\lambda[\bar{\lambda}]>0$ and $\bar{\alpha}\in\supp(\gamma_{\lambda,\bar{\lambda}})$.

\begin{definition}\label{def: incomplete information blocking}
Fix an outcome with incomplete information $(\Lambda,p, q, \mu, \gamma)$. An \textbf{incomplete-information blocking pair} exists if there are types $t, t'\in\{\theta, \tau\}$; labels $\lambda, \lambda'\in \Lambda$ with $q_\lambda[t]>0$ and $q_{\lambda'}[t']>0$; current positions $(\bar{\lambda},\bar{\alpha})$ of label $\lambda$ and $(\bar{\lambda}',\bar{\alpha}')$ of label $\lambda'$; and an agreement $\hat{\alpha}\in\A$ such that for any deviation plans $(\lambda, D, d)$ and $(\lambda',D', d')$ satisfying $t\in D$, $t'\in D'$, $d(t)=f$, and $d'(t')=f$, we have
\begin{itemize}
\setlength{\itemsep}{0pt}
\item[(i)] $f \in \argmax_{g}\,\mathbb{E}_{\hat{\alpha}}[\mathbb{E}_{q_{\lambda'}} [u_t(g(a), d'(\cdot)(a'), \cdot)\,|\,D']]$ and\\
$f \in \argmax_{g}\,\mathbb{E}_{\rho(\hat{\alpha})}[\mathbb{E}_{q_\lambda} [u_{t'}(g(a), d(\cdot)(a'), \cdot)\,|\,D]]$;
\item[(ii)] $\mathbb{E}_{\hat{\alpha}}[\mathbb{E}_{q_{\lambda'}} [u_t(a, d'(\cdot)(a'), \cdot)\,|\,D']] > \mathbb{E}_{\bar{\alpha}}[\mathbb{E}_{q_{\bar{\lambda}}} [u_t(a, a', \cdot)]]$ and\\
$\mathbb{E}_{\rho(\hat{\alpha})}[\mathbb{E}_{q_{\lambda}} [u_{t'}(a, d(\cdot)(a'), \cdot)\,|\,D]] > \mathbb{E}_{\bar{\alpha}'}[\mathbb{E}_{q_{\bar{\lambda}'}} [u_{t'}(a, a', \cdot)]]$.
\end{itemize}
\end{definition}

In the definition above, for an agent of type $t\in\{\theta,\tau\}$, a pair of action recommendations $(a,a')$, and a deviation plan $(\lambda',D', d')$ of the deviating partner, the conditional expected utility $\mathbb{E}_{q_{\lambda'}} [u_t(g(a), d'(\cdot)(a'), \cdot)\,|\,D']$ is evaluated using the probability distribution $q_{\lambda'}$ conditional on the partner's type belonging to $D'$. If $D'$ is a singleton, then the expectation is degenerate. 

Definition \ref{def: incomplete information blocking} describes a situation in which a pair of agents, despite observing only each other's labels, can still reach an agreement and carry out a mutually beneficial deviation.\footnote{Our approach is akin to the cheap talk paradigm, in which information is conveyed through endogenous behavior. This is fundamentally different from models in which agents incur costs to signal their types or increase the likelihood of being identified.} In particular, a blocking pair consists of a type-$t$ agent with label $\lambda$ (i.e., a type-$t_\lambda$ agent) and a type-$t'$ agent with label $\lambda'$ (i.e., a type-$t'_{\lambda'}$ agent). We call them the \textbf{targeted} agents. Provided that the targeted type-$t'_{\lambda'}$ agent participates and follows $\hat{\alpha}$, the type-$t_\lambda$ agent also finds it optimal to follow $\hat{\alpha}$ and thereby obtains a strict utility improvement, regardless of whether or how nontargeted agents with label $\lambda'$ participate in the deviation. The same argument applies symmetrically to the targeted type-$t'_{\lambda'}$ agent. Thus, each targeted agent's incentive to deviate is guaranteed conditional on the participation and obedience of her targeted partner.\footnote{Consider the following statement a type-$t_\lambda$ agent could make to an agent with label $\lambda'$: ``I am of type $t$, and I promise to participate in the deviation and follow the recommendation of $\hat{\alpha}$ for me. I ask you to join the deviation and follow the recommendation of $\hat{\alpha}$ for you if you are of type $t'$. Why should you follow my advice? Because as long as I fulfill my promise, following the recommendation of $\hat{\alpha}$ is optimal for you, which strictly improves your expected utility. Why should you trust my promise? Because as long as you follow my advice, following the recommendation of $\hat{\alpha}$ is optimal for me, which strictly improves my expected utility.''} 


\begin{remark}
    In Definition \ref{def: incomplete information blocking}, a deviating agent relies on the rationality of her targeted partner, conditional on her own participation. While it is possible to further account for the rationality of nontargeted types, we adopt a more conservative approach for two reasons. First, rational behavior of nontargeted agents may reveal additional information in a deviation and may induce the deviating partner to adjust her behavior after the rematch.\footnote{This occurs, for example, when the deviation plan $(\lambda,D, d)$ satisfies $D=\{\theta,\tau\}$ and $d(\theta)\neq d(\tau)$.} Accounting for this possibility would require us to take a stance on how agents anticipate and respond to the resulting inferences and behavioral adjustments; see a related discussion in \citet{liu2020AER}. Our definition avoids this issue because the responses of nontargeted agents---whether rational or irrational---do not affect the decisions of the targeted agents. Second, a more demanding criterion for blocking yields a weaker notion of stability, thereby strengthening the positive result in Proposition \ref{prop: parochial efficient with ii}. At the same time, the negative result in Proposition \ref{prop: morally constrained dominated} does not rely on irrational behavior by any nontargeted type; see footnotes \ref{ft: rational play} and \ref{ft: rational play 2}.
\end{remark}

We use an example to illustrate the notion of an incomplete-information blocking pair.

\begin{example}\label{eg: incomplete-information blocking}
Consider the prisoners' dilemma material game as follows:
    
\begin{center}
    \begin{tabular}{cc|c|c|}
       & \multicolumn{1}{c}{} & \multicolumn{1}{c}{$x$}  & \multicolumn{1}{c}{$y$} \\\cline{3-4}
       & $x$ & $\  4,4 \ $ & $\ 1,5 \ $ \\\cline{3-4}
      & $y$ & $\  5,1 \ $ & $\ 3,3 \ $ \\\cline{3-4}
    \end{tabular}
\end{center}

Suppose that type-$\theta$ agents have efficient preferences $u_\theta(a, a', t)=\pi(a, a')+\pi(a', a)$, while type-$\tau$ agents are selfish $u_\tau(a, a', t)=\pi(a, a')$. Consider a population state $(\theta,\tau,\varepsilon)$ and an outcome $(\Lambda, p, q, \mu, \gamma)$ as follows. Each type constitutes one half of the population, so $\varepsilon=1/2$. The matching profile $(\Lambda, p, q, \mu)$ satisfies $\Lambda=\{\lambda\}$, $p[\lambda]=1$, $q_\lambda[\theta]=q_\lambda[\tau]=1/2$, and $\mu_{\lambda}[\lambda]=1$. The agreement profile $\gamma$ satisfies $\gamma_{\lambda,\lambda}[\alpha]=1$, where $\alpha[y,y]=1$. This outcome is depicted in Figure \ref{eg pic: incomplete-information blocking} below.

\begin{figure}[!ht]
    \centering
    \begin{tikzpicture}[scale=0.85]
    \draw[black!80] (-4,-0.4) rectangle +(8,0.8);
    
    \fill[red, fill opacity=0.2] (0,-0.4) rectangle +(4,0.8);
    
    \fill[blue, fill opacity=0.2] (-4,-0.4) rectangle +(4,0.8);
    
    \node at (0,0) {$\lambda$};

    \draw [decorate,decoration={brace,amplitude=5pt,mirror},xshift=0pt,yshift=-8pt]
    (0,-0.4) -- (4,-0.4) node [midway,yshift=-12pt, xshift=0pt] 
    {$\tau$};
    
    \draw [decorate,decoration={brace,amplitude=5pt,mirror},xshift=0pt,yshift=-8pt]
    (-4,-0.4) -- (0,-0.4) node [midway,yshift=-12pt, xshift=0pt] 
    {$\theta$};
    
    \draw[black!40, rounded corners, dashed] (-4.1,-0.5) rectangle (4.1,0.5);
    \fill[pattern=north east lines, pattern color=black!30,rounded corners] (-4.1,-0.5) rectangle (4.1,0.5);
    
    \draw [<-] (+0.2,0.6) to [out=30, in=270] (+0.4,1);
    \draw [-] (+0.4,1) to [out=90, in=0] (0,1.4);
    \draw [-] (0,1.4) to [out=180, in=90] (-0.4,1);
    \draw [->] (-0.4,1) to [out=270, in=150] (-0.2,0.6);
    \node at (0,1.4) [above=] {$\alpha$};

    \end{tikzpicture}
    \caption{The outcome in Example \ref{eg: incomplete-information blocking}.}
    \label{eg pic: incomplete-information blocking}
\end{figure}

To show that an incomplete-information blocking pair exists, consider two type-$\theta_\lambda$ agents who target one another and propose the efficient agreement $\hat{\alpha}$, where $\hat{\alpha}[x, x]=1$. By symmetry, it suffices to verify the incentives of one of these agents. Consider any deviation plan $(\lambda, D, d)$ of her deviating partner satisfying $\theta\in D$ and $d(\theta)=f$; that is, the targeted type-$\theta_\lambda$ partner participates in the deviation and follows $\hat{\alpha}$. We show that following $\hat{\alpha}$ is optimal for the agent and makes her strictly better off. There are two cases to consider:

\begin{itemize}
    \item Suppose $\tau\notin D$. Since $x$ is a best response against $x$ for a type-$\theta$ agent, following $\hat{\alpha}$ is optimal. Moreover, following $\hat{\alpha}$ yields a utility of $8$ to the deviating type-$\theta$ agent, exceeding her status quo expected utility of $6$.
    \item Suppose $\tau\in D$. Because $\hat{\alpha}$ recommends $x$ with
    probability one, only the action $d(\tau)(x)$ of the deviating partner matters. If $d(\tau)(x)=x$, the partner plays $x$ regardless of her type. Following $\hat{\alpha}$ is therefore optimal and gives the type-$\theta$ agent expected utility $8>6$. If $d(\tau)(x)=y$, following $\hat{\alpha}$ is still optimal because $8(1/2)+6(1/2)=7>6=6(1/2)+6(1/2)$, which also makes her strictly better off.
\end{itemize}

Therefore, conditional on a targeted type-$\theta_\lambda$ partner participating in the deviation and following $\hat{\alpha}$, a type-$\theta_\lambda$ agent also finds it optimal to follow $\hat{\alpha}$ and obtains a strict utility improvement. This conclusion remains valid even if nontargeted type-$\tau_\lambda$ agents join the deviation and behave arbitrarily. Hence, an incomplete-information blocking pair exists.
\end{example}

We extend the notion of stable outcome to the case of incomplete information.

\begin{definition}\label{def: stable outcome incomplete info}
An outcome with incomplete information $(\Lambda, p, q, \mu, \gamma)$ is \textbf{incomplete-information stable} if it satisfies:
\begin{itemize}
\item[\rm{(i)}] (Internal stability) $\gamma$ is a correlated equilibrium profile with respect to $\mu$;
\item[\rm{(ii)}] (External stability) There is no incomplete-information blocking pair.
\end{itemize}
\end{definition}

When $\Lambda=\{\lambda,\lambda'\}$, $p[\lambda]=1-\varepsilon$, $p[\lambda']=\varepsilon$, and $q_\lambda[\theta]=q_{\lambda'}[\tau]=1$, agents' types are fully revealed, and Definitions \ref{def: incomplete-information correlated equilibrium profile} and \ref{def: incomplete information blocking} then coincide with Definitions \ref{def: correlated equilibrium profile} and \ref{def: complete information blocking}, respectively. Hence, incomplete-information stability reduces to complete-information stability. The existence of an incomplete-information stable outcome is therefore guaranteed. An outcome that is stable under incomplete information, however, need not remain stable if agents' preference types become fully observable, as full disclosure may create additional blocking opportunities. As highlighted in Remark \ref{rmk: endogenous info}, information in a stable outcome is \emph{endogenous}: the degree to which preference types are revealed depends on agents' matching patterns and behavior.

\subsection{Evolutionary Dominance with Incomplete Information}\label{sec: evo dominance with incomplete info}

Given an incomplete-information stable outcome $(\Lambda, p, q, \mu, \gamma)$ in population state $(\theta,\tau,\varepsilon)$, the average material payoffs for agents of type $\theta$ and type $\tau$ are given by
\begin{align*}
G_\theta(\Lambda, p, q, \mu,\gamma)&=\frac{1}{1-\varepsilon}\sum_{\lambda\in\Lambda}\left\{\sum_{\lambda'\in\Lambda}\left[\int_{\A}\mathbb{E}_{\alpha}[\pi(a,a')]\mathop{d \gamma_{\lambda,\lambda'}[\alpha]}\right] \mu_\lambda[\lambda']\right\} p[\lambda]q_{\lambda}[\theta],\\
G_\tau(\Lambda, p, q, \mu,\gamma)&=\frac{1}{\varepsilon}\sum_{\lambda\in\Lambda}\left\{\sum_{\lambda'\in\Lambda}\left[\int_{\A}\mathbb{E}_{\alpha}[\pi(a,a')]\mathop{d \gamma_{\lambda,\lambda'}[\alpha]}\right] \mu_\lambda[\lambda']\right\} p[\lambda]q_{\lambda}[\tau].
\end{align*}

Our notions of evolutionary dominance can be naturally extended to incorporate incomplete information.

\begin{definition}\label{def: evo dominance with incomplete info}
Fix a material game $\G$. With incomplete information,
\begin{itemize}
    \item[(i)] A preference type $\theta \in \Theta$ is \textbf{evolutionarily dominant} if for every $\tau \in \Theta$ and every $\varepsilon\in(0,1)$, every incomplete-information stable outcome $(\Lambda,p,q,\mu,\gamma)$ in the population state $(\theta,\tau,\varepsilon)$ satisfies $G_\theta(\Lambda,p,q,\mu,\gamma)\geq G_\tau(\Lambda,p,q,\mu,\gamma)$. 
    \item[(ii)] A preference type $\theta\in\Theta$ is \textbf{evolutionarily dominated} if there exists another type $\tau\in\Theta$ such that for every $\varepsilon\in(0,1)$, every incomplete-information stable outcome $(\Lambda,p,q,\mu,\gamma)$ in the population state $(\theta,\tau,\varepsilon)$ satisfies $G_\theta(\Lambda,p,q,\mu,\gamma)\leq G_\tau(\Lambda,p,q,\mu,\gamma)$, with strict inequality for some incomplete-information stable outcome.
\end{itemize}
\end{definition}

A morally constrained selfish type is evolutionarily dominant under complete information (Proposition \ref{prop: morally constrained selfish}) because agents of such a type can secure a material payoff of $m/2$ by rematching with one another to play a fair and efficient agreement. This mechanism, however, relies critically on complete information. When morally constrained selfish agents can no longer identify others of their own type, they may be reluctant to undertake such deviations, causing the type to lose its evolutionary advantage. The following example illustrates this possibility.

\begin{example}\label{eg: morally constrained dominated}
Consider again the material game from Example \ref{eg: complete-info stability}, reproduced below:
\begin{center}
  \begin{tabular}{c|c|c|}
    \multicolumn{1}{c}{}    & \multicolumn{1}{c}{$x$}  & \multicolumn{1}{c}{$y$} \\\cline{2-3}
      $x$ & $\  1,1 \ $ & $\ 4,2 \ $  \\\cline{2-3}
       $y$ & $\  2,4 \ $ & $\ 1,1 \ $  \\\cline{2-3}
    \end{tabular}
\end{center}

Let $\theta$ denote a morally constrained selfish type. Let $\tau$ denote a type with efficient preferences when matched with its own type but for which $x$ is strictly dominant otherwise:
\begin{equation*}
   u_\tau(a, a',t) = \begin{cases}
    \pi(a, a')+\pi(a', a) &\text{if $t=\tau$,}\\
     8\cdot\mathbf{1}_{\{a=x\}} &\text{if $t\neq \tau$.}
    \end{cases}
\end{equation*}

Now consider a population state $(\theta,\tau,\varepsilon)$ and an outcome $(\Lambda, p, q, \mu, \gamma)$ as follows. Assume that the proportion of type-$\tau$ agents satisfies $\varepsilon>3/4$. The matching profile $(\Lambda, p, q, \mu)$ satisfies $\Lambda=\{\lambda,\lambda'\}$, $p[\lambda]=p[\lambda']=1/2$, $q_{\lambda}[\theta]=2(1-\varepsilon)<1/2$, $q_{\lambda'}[\tau]=1$, and $\mu_\lambda[\lambda']=\mu_{\lambda'}[\lambda]=1$. The agreement profile $\gamma$ satisfies $\gamma_{\lambda,\lambda'}[\alpha]=1$, where $\alpha[y,x]=1$. This outcome is depicted in Figure \ref{eg pic: morally constrained dominated}.

\begin{figure}[!ht]
    \centering
    \begin{tikzpicture}[scale=0.85]
    \draw[black!80] (-4,-0.4) rectangle +(8,0.8);
    
    \fill[red, fill opacity=0.2] (-2.5,-0.4) rectangle +(6.5,0.8);
    \node at (2,0) {$\lambda'$};
    
    \fill[blue, fill opacity=0.2] (-4,-0.4) rectangle +(1.5,0.8);
    
    \node at (-2,0) {$\lambda$};

    \draw [decorate,decoration={brace,amplitude=5pt,mirror},xshift=0pt,yshift=-8pt]
    (-2.5,-0.4) -- (4,-0.4) node [midway,yshift=-12pt, xshift=0pt] 
    {$\tau$};
    
    \draw [decorate,decoration={brace,amplitude=5pt,mirror},xshift=0pt,yshift=-8pt]
    (-4,-0.4) -- (-2.5,-0.4) node [midway,yshift=-12pt, xshift=0pt] 
    {$\theta$};
    

    \draw[black!40, rounded corners, dashed] (-4.1,-0.5) rectangle (0,0.5);
    \fill[pattern=north east lines, pattern color=black!30,rounded corners] (-4.1,-0.5) rectangle (0,0.5);

    \draw[black!40, rounded corners, dashed] (0,-0.5) rectangle (4.1,0.5);
    \fill[pattern=dots, pattern color=black!30,rounded corners] (0,-0.5) rectangle (4.1,0.5);
    
    \draw [<-] (2,0.6) to [out=110, in=0] (0,1.4);
    \draw [->] (0,1.4) to [out=180, in=70] (-2,0.6);
    \node at (0,1.4) [above=] {$\alpha$};
    \end{tikzpicture}
    \caption{The outcome in Example \ref{eg: morally constrained dominated}.}
    \label{eg pic: morally constrained dominated}
\end{figure}

We formally verify in Appendix \ref{proof of eg: morally constrained dominated} that $(\Lambda,p,q,\mu,\gamma)$ is incomplete-information stable. For intuition, suppose type-$\theta_\lambda$ agents target one another and propose, for example, a fair and efficient agreement $\hat{\alpha}$ satisfying $\hat{\alpha}[x,y]=\hat{\alpha}[y,x]=1/2$. Although both sides would follow $\hat{\alpha}$  and gain under complete information, nontargeted type-$\tau_\lambda$ agents may join the deviation and play $x$ under incomplete information, leaving the targeted agents with expected utility $((2+4)/2)q_\lambda[\theta]+((2+0)/2)(1-q_\lambda[\theta]) =1+2q_\lambda[\theta]<2$, below their status quo utility. Thus, the deviation is not viable under incomplete information. Notably, although $\alpha$ is efficient, type-$\theta$ agents earn a strictly lower average material payoff than type-$\tau$ agents in this incomplete-information stable outcome.
\end{example}

Example \ref{eg: morally constrained dominated} constructs a particular incomplete-information stable outcome in which the morally constrained selfish type earns a lower average material payoff than the competing type. The following proposition shows that the underlying vulnerability is more general: whenever efficient play is unique up to role reversal and yields unequal material payoffs, a morally constrained selfish type is evolutionarily dominated under incomplete information.\footnote{This class encompasses many familiar games in which efficient play requires agents to assume different roles, with one role yielding a larger material payoff than the other. Canonical examples include the Battle of the Sexes \citep{LuceRaiffa1957} (with a suitably redefined action set), the Leader game \citep{Rapoport1967}, and the two-player volunteer's dilemma \citep{Diekmann1985}. Even outside this class, simple examples can be constructed in which a morally constrained selfish type fails to be evolutionarily dominant.}

\begin{proposition}\label{prop: morally constrained dominated}
With incomplete information, if $\E=\{(a^*,a^\dagger),(a^\dagger,a^*)\}$ and $\pi(a^*,a^\dagger)\neq\pi(a^\dagger,a^*)$, then a morally constrained selfish type is evolutionarily dominated.
\end{proposition}

The proof constructs a competing type with two complementary properties. First, regardless of its population share, this type earns an average material payoff of at least $m/2$ in every incomplete-information stable outcome. This ensures that it always weakly outperforms the morally constrained selfish type (Lemma \ref{lemma: prop 5 weak}). Second, its behavior sometimes deters morally constrained selfish agents from escaping the disadvantaged side of efficient play, allowing the payoff inequality to be strict in some stable outcome (Lemma \ref{lemma: prop 5 strict}). Together, these properties establish evolutionary domination.

In contrast, the next proposition shows that a parochial efficient type stands out even with incomplete information.

\begin{proposition}\label{prop: parochial efficient with ii}
With incomplete information, for any material game $\G$, a parochial efficient type is evolutionarily dominant.
\end{proposition}

To understand the sharp contrast between Propositions \ref{prop: morally constrained dominated} and \ref{prop: parochial efficient with ii}, it is helpful to consider the underlying logic of Example \ref{eg: morally constrained dominated}. There, morally constrained selfish agents cannot escape an unfavorable position through pairwise deviations because nontargeted type-$\tau_\lambda$ agents may join and behave in ways that make the proposed rematch unprofitable. Parochial efficient agents, by contrast, place first priority on matching with their own type, so the behavior of nontargeted type-$\tau_\lambda$ agents cannot eliminate the gains from rematching. They can therefore break away from the same disadvantageous position.

\begin{example}[Example \ref{eg: morally constrained dominated} revisited]
    Consider the population state and outcome from Example \ref{eg: morally constrained dominated}, except that $\theta$ now denotes a parochial efficient type. Type-$\theta_\lambda$ agents derive utility $0$ in the status quo. Thus, two type-$\theta_\lambda$ agents can target one another and propose the efficient agreement $\hat{\alpha}$ from Example \ref{eg: morally constrained dominated}. If the nontargeted type-$\tau_\lambda$ agents do not participate, each targeted agent finds it optimal to follow $\hat{\alpha}$ and obtains expected utility $6>0$. If they also participate, following $\hat{\alpha}$ remains optimal for each targeted agent and yields expected utility $6q_{\lambda}[\theta]+0(1-q_{\lambda}[\theta])>0$, regardless of how the nontargeted agents behave. Thus, the proposed deviation constitutes an incomplete-information blocking pair.
\end{example}


The proof of Proposition \ref{prop: parochial efficient with ii} shows how parochial efficient agents overcome informational frictions. Specifically, when a parochial efficient type is present, incomplete-information stable outcomes satisfy two key properties (see Lemmas \ref{lemma: perfect assortativity} and \ref{lemma: efficient play}). First, matching is perfectly assortative by label: matched partners always carry the same label. This, on average, limits the competing type's ability to gain from any material-payoff sacrifice made by the parochial efficient type. Second, for every label $\lambda$ such that $q_\lambda[\theta]>0$, agents carrying that label interact efficiently with one another. Together, these properties ensure that a parochial efficient type attains an average material payoff at least as high as that of any competing type.

\section{Polymorphism}\label{sec: polymorphism}
Thus far, we have focused on evolutionary dominance of a single type. In this section, we extend the framework to polymorphic populations, in which several preference types coexist. For simplicity, we maintain complete information in this section; the corresponding extension under incomplete information is conceptually similar.

Let $\nu\in\Delta(\Theta)$ be a \textbf{population distribution} with finite support $\Theta_\nu=\supp(\nu)$. For each type $t\in\Theta_\nu$, the mass of type-$t$ agents in the population is denoted by $\nu[t]>0$. An outcome $(\mu,\gamma)$ is defined as in the two-type case. The matching profile is now a vector $\mu=(\mu_t)_{t\in\Theta_\nu}$, where $\mu_t\in\Delta(\Theta_\nu)$ gives the distribution of partner types for type-$t$ agents. In particular, $\mu_t[t']$ is the fraction of type-$t$ agents matched with type-$t'$ agents. The matching profile satisfies the consistency condition
\[
\nu[t]\mu_t[t']=\nu[t']\mu_{t'}[t]
\qquad\text{for all $t,t'\in\Theta_\nu$}.
\]
For each $t,t'\in\Theta_\nu$, let $\gamma_{t,t'}\in\Delta(\A)$ denote the conditional distribution of agreements reached in matches between type-$t$ and type-$t'$ agents. As before, the agreement profile $\gamma=(\gamma_{t,t'})$ satisfies the exchangeability condition.

The definitions of a complete-information blocking pair and a complete-information stable outcome under a population distribution $\nu$ extend directly by replacing $\{\theta,\tau\}$ with $\Theta_\nu$ in Definitions \ref{def: complete information blocking} and \ref{def: stable outcome}. The resulting environment is a one-sided matching market with contracts and a continuum of agents. An important recent literature studies existence and structural properties of stable outcomes in related large matching markets \citep{CheKimKojima2019,Greinecker2021,Jagadeesan2024, carmona2024}. Adapting arguments from this literature, we establish the following existence result.

\begin{proposition}\label{prop: existence}
For any material game $\G$ and any population distribution $\nu$, a complete-information stable outcome exists.
\end{proposition}

Given a complete-information stable outcome $(\mu,\gamma)$ under a population distribution $\nu$, for each $t\in \Theta_\nu$, the average material payoff earned by type-$t$ agents is given by 
\begin{align*}
G_t(\mu,\gamma)&=\sum_{t'\in \Theta_{\nu}}\left[\int_{\mathcal{A}}\mathbb{E}_{\alpha}[\pi(a, a')]\mathop{d\gamma_{t,t'}[\alpha]}\right]\mu_{t}[t'].
\end{align*}

We next follow \citet{Dekeletal2007RESTUD} and ask whether a population distribution can resist invasion by any competing preference type. We assume that mutations are sufficiently rare and happen one at a time, with the population able to fully adjust before the next mutation occurs \citep{Weibull1995}. Accordingly, we treat the population distribution as incumbent and restrict the mutant to a sufficiently small population share. This leads to the following local notion of evolutionary dominance.

\begin{definition}\label{def: dominance of population}
    A population distribution $\nu$ is \textbf{locally evolutionarily dominant} if there exists $\bar{\varepsilon}>0$ such that for every mutant type $t'\in\Theta$, $\varepsilon\in(0,\bar{\varepsilon})$, and complete-information stable outcome $(\tilde{\mu}, \tilde{\gamma})$ under the post-entry population distribution $\tilde{\nu}=(1-\varepsilon)\nu+\varepsilon \delta_{t'}$, we have $G_t(\tilde{\mu}, \tilde{\gamma})\geq G_{t'}(\tilde{\mu}, \tilde{\gamma})$ for every $t\in \Theta_\nu$.\footnote{We write $\delta_{t'}\in\Delta(\Theta)$ for the Dirac measure that assigns probability one to type $t'$.}
\end{definition}

Thus, whenever the mutant's population share lies below a common threshold, every incumbent type must earn at least as much as the mutant in every post-entry stable outcome. No sufficiently small invasion can therefore displace any type in the incumbent population. Under this definition, the mutant type may already belong to the incumbent support, in which case the perturbation changes only its population share.

We now derive necessary conditions that describe crucial properties of population distributions that are locally evolutionarily dominant.

\begin{proposition} \label{prop: necessary condition for polymorphism}
Suppose that the population distribution $\nu$ is locally evolutionarily dominant. Then every complete-information stable outcome $(\mu, \gamma)$ under $\nu$ satisfies the following:
\begin{itemize}
    \item[(i)] $G_t(\mu, \gamma)=G_{t'}(\mu, \gamma)$ for all $t,t' \in \Theta_\nu$.
    \item[(ii)] For any $t,t'\in \Theta_\nu$ such that $\mu_{t}[t']>0$, every agreement $\alpha\in \supp(\gamma_{t, t'})$ is efficient. Moreover, either $t=t'$ or $\mathbb{E}_\alpha[\pi(a, a')]=\mathbb{E}_{\rho(\alpha)}[\pi(a, a')]$.
\end{itemize}
\end{proposition}

Part (i) of Proposition \ref{prop: necessary condition for polymorphism} means that a locally evolutionarily dominant $\nu$ should itself be \textit{balanced}: all types in its support must earn the same average material payoff. Otherwise, natural selection would alter their relative frequencies even in the absence of mutations. Whereas \citet{Dekeletal2007RESTUD} impose balancedness as part of their definition of evolutionary stability for a polymorphic population, here it follows from local evolutionary dominance.

Part (ii) requires efficient play in every complete-information stable outcome. If some incumbents play inefficiently, a parochial efficient mutant can separate from the incumbent population and play efficiently with its own type, resulting in a post-entry stable outcome in which the mutant outperforms at least one incumbent type. Part (ii) also imposes a form of payoff symmetry on cross-type matches: whenever two distinct types are matched, the two sides must receive the same expected material payoff.

While the necessary conditions in Proposition \ref{prop: necessary condition for polymorphism} may appear demanding, the following result provides a sufficient condition.

\begin{proposition} \label{prop: stable polymorphism}
If $\Theta_\nu$ consists of morally constrained selfish and parochial efficient types, then $\nu$ is locally evolutionarily dominant. 
\end{proposition}

The result follows from the logic of Propositions \ref{prop: morally constrained selfish} and \ref{prop: parochial efficient}, and we therefore omit a formal proof. In any stable outcome following the entry of an arbitrary mutant, all morally constrained selfish incumbents earn at least $m/2$, while parochial efficient incumbents match with their own type and play efficiently, earning $m/2$ on average. Because the aggregate material payoff of the population is bounded above by $m/2$, the mutant earns at most this bound on average. Thus, all incumbent types weakly outperform the mutant.

It is natural to ask whether a locally evolutionarily dominant population can contain preference types beyond the two kinds considered above. Of particular interest are mutually \emph{heterophilic} types---distinct types that prefer to match with one another. Such preferences may arise when social identities or observable characteristics shape partner choice and behavior. The following example shows that heterophily can indeed be sustained.

\begin{example}
Consider again the prisoners' dilemma material game in Example \ref{eg: incomplete-information blocking}, which is reproduced below:
\begin{center}
  \begin{tabular}{c|c|c|}
    \multicolumn{1}{c}{}    & \multicolumn{1}{c}{$x$}  & \multicolumn{1}{c}{$y$} \\\cline{2-3}
      $x$ & $\  4,4 \ $ & $\ 1,5 \ $  \\\cline{2-3}
       $y$ & $\  5,1 \ $ & $\ 3,3 \ $  \\\cline{2-3}
    \end{tabular}
\end{center}

Let $\nu$ denote a population distribution that contains two types $\Theta_\nu=\{\theta,\tau\}$ and $\nu[\theta]=\nu[\tau]=1/2$. The utility functions of $\theta$ and $\tau$ are given by, respectively,
\begin{equation*}
   u_\theta(a, a',t) = \pi(a,a') + \pi(a',a)\cdot\mathbf{1}_{\{t=\tau\}} \quad\text{and}\quad
    u_{\tau}(a,a',t) = \pi(a,a') + \pi(a',a)\cdot\mathbf{1}_{\{t=\theta\}}
\end{equation*}

We now show that $\nu$ is locally evolutionarily dominant. Consider a mutant type $t'$, an $\varepsilon>0$ sufficiently small, and an arbitrary complete-information stable outcome $(\tilde{\mu},\tilde{\gamma})$ under the post-entry population $\tilde{\nu}=(1-\varepsilon)\nu+\varepsilon \delta_{t'}$. There are two cases:
\begin{itemize}
    \item If $t'\notin\Theta_\nu$, then we must have $\tilde{\mu}_\theta[\tau]=\tilde{\mu}_{\tau}[\theta]=1$ and $\tilde{\gamma}_{\theta,\tau}[\hat{\alpha}]=1$, where $\hat{\alpha}[x, x]=1$. It follows that $G_{\theta}(\tilde{\mu},\tilde{\gamma})=G_{\tau}(\tilde{\mu},\tilde{\gamma})=4\geq G_{t'}(\tilde{\mu},\tilde{\gamma})$.
    \item If $t'\in\Theta_\nu$, suppose $t'=\theta$ without loss. Then all type-$\tau$ agents match with type-$\theta$ agents and play $(x,x)$, while excess type-$\theta$ agents must match among themselves. Therefore, $\tilde{\mu}_\theta[\theta]=2\varepsilon/(1+\varepsilon)$, $\tilde{\mu}_\theta[\tau]=(1-\varepsilon)/(1+\varepsilon)$, and $\tilde{\mu}_{\tau}[\theta]=1$; moreover, $\tilde{\gamma}_{\theta,\tau}[\hat{\alpha}]=1$ and $\tilde{\gamma}_{\theta,\theta}[\tilde{\alpha}]=1$, where $\tilde{\alpha}[y, y]=1$. Consequently, $G_{\theta}(\tilde{\mu},\tilde{\gamma})<4=G_{\tau}(\tilde{\mu},\tilde{\gamma})$. Both incumbent types weakly outperform the entrant.
\end{itemize}
Therefore, the population distribution $\nu$ that consists of two heterophilic types is locally evolutionarily dominant. Interestingly, although both types would play inefficiently with same-type partners, this does not destabilize the population because such matches do not occur at the balanced composition. 
\end{example}

\section{Concluding Remarks}

This paper studies preference evolution under endogenous matching by combining stable partner choice with equilibrium play. Evolutionary success requires agents both to sustain efficient play and to protect themselves from exploitation. We identify two classes of preference types that combine these forces: morally constrained selfish types and parochial efficient types. Both classes are evolutionarily dominant under complete information. Under incomplete information, however, morally constrained selfish types can be dominated, whereas parochial efficient types remain dominant. By prioritizing same-type matching, parochialism acts as a form of group-level selfishness and enables agents to overcome the informational frictions that impede assortative matching.

Although \citet{CharnessRabin2002QJE} and \citet{engelmann2004inequality} find behavior consistent with efficiency concerns, experimental support for an unconditional preference for efficiency remains inconclusive. Our analysis suggests that efficient behavior may reflect more structured motives: agents may either maximize their own material payoffs subject to an efficiency constraint or value efficiency only when interacting with their own type. Standard experimental designs based on exogenous matching are therefore not well suited to detect and distinguish these motives. A more informative design would allow subjects to communicate their intended play and form pairs by mutual consent.\footnote{An experimental literature beginning with \citet{EhrhartKeser1999} and \citet{Hauk2001} studies behavior under endogenous partner choice. A common finding is that partner choice promotes cooperation and coordination on efficient outcomes by allowing like-minded subjects to associate. Related experiments find similar effects when subjects are grouped assortatively by elicited cooperativeness \citep{Coricelli2004,Burlando2005,deOliveira2015}.} Once matching patterns and play have stabilized, researchers could, for example, elicit subjects' other-regarding preferences separately toward current partners and nonpartners. Greater weight on efficiency in choices involving current partners than in otherwise identical choices involving nonpartners would provide evidence consistent with parochial efficiency.

Several important questions remain for future research. First, a dynamic model could replace our reduced-form notion of stability with history-dependent matching and play. The analysis of repeated coalitional behavior by \citet{ali2025} suggests that agents with perfectly aligned preferences attain a jointly optimal outcome in every dynamically stable process. Because the interests of two parochial efficient agents are indeed perfectly aligned, evolutionary dominance of such types may extend to a genuinely dynamic setting. Second, preferences and institutions may coevolve. Institutions shape incentives for matching and play through taxes and subsidies \citep{HillerToure2021JDE}, property-rights protection \citep{BisinVerdier2021BC}, religious infrastructure \citep{Bisinetal2021WP}, and integration policies \citep{wu2017political}. Extending our framework along the lines of \citet{bisin2024joint} would allow a joint analysis of the evolution of preferences and institutions.

\renewcommand*{\bibfont}{\small\linespread{0.97}\selectfont}

\bibliographystyle{ecta}
\bibliography{bib}

\begin{appendix}

\section{Omitted Proofs}

\subsection{Proofs for Section \ref{sec: preference evolution complete info}: Complete Information}

\subsubsection{Proof of Proposition \ref{prop: morally constrained selfish}} \label{proof: morally constrained selfish}
We first establish the following lemma, which guarantees the existence of an efficient correlated equilibrium between two morally constrained selfish agents.

\begin{lemma} \label{lemma: efficient CE}
Let $\theta$ denote a morally constrained selfish type. There exists an agreement $\tilde{\alpha}\in\CE_{\theta,\theta}$ such that $\tilde{\alpha}[\E]=1$.
\end{lemma}

\begin{proof}
We choose the following normalization of a morally constrained selfish type,
\[
u_\theta(a,a',\theta)=
\begin{cases}
\pi(a,a'), & \text{if }(a,a')\in\E,\\
0, & \text{otherwise}.
\end{cases}
\]
Let
\[
F=\{a\in A:(a,a')\in\E \text{ for some } a'\in A\}
\]
and consider the finite symmetric game on $F$ with utility function $\tilde{u}(a,a')=u_\theta(a,a',\theta)$ for $(a,a')\in F^2$. Because the game is finite and symmetric, it admits a symmetric Nash equilibrium. Let $x\in\Delta(F)$ be one; that is, for every $a\in\supp(x)$,
\begin{equation}\label{eq: NE condition}
    \sum_{a'\in F}\tilde{u}(a,a') x[a']\geq \sum_{a'\in F}\tilde{u}(\hat{a},a') x[a'] \ \ \text{for all $\hat{a}\in F$}.
\end{equation}

For each $a\in F$, define
\[
H(a)=\{a'\in F:(a,a')\in\E\},
\]
and let
\[
P=\sum_{(a,a')\in\E}x[a]x[a'].
\]
We first show that $P>0$. Suppose instead that $P=0$. Then every pair of actions in $\supp(x)$ is inefficient, so every supported action yields expected utility zero. Choose any $a\in\supp(x)$. Since $a\in F$, there exists $\hat{a}\in F$ such that $(a,\hat{a})\in\E$. Because $\E$ is symmetric, $(\hat{a},a)\in\E$. Deviating to $\hat{a}$ therefore yields expected payoff at least
\[
\pi(\hat{a},a)x[a]>0,
\]
contradicting the optimality of the supported action $a$. Hence, $P>0$.

Define an agreement $\tilde{\alpha}$ by conditioning the pairwise product of $x$ on $\E$:
\[
\tilde{\alpha}[a,a'] =
\begin{cases}
\dfrac{x[a]x[a']}{P}, & \text{if }(a,a')\in\E,\\
0, & \text{otherwise}.
\end{cases}
\]
By construction, $\tilde{\alpha}$ is symmetric and satisfies $\tilde{\alpha}[\E]=1$.

It remains to verify the incentive constraints of a correlated equilibrium. Fix a recommendation $a$ that occurs with positive probability and a deviation $\hat{a}\in A$. Then $x[a]>0$. If $\hat{a}\notin F$, then $u_\theta(\hat{a},a',\theta)=0$ for every $a'\in A$. Because $a$ yields strictly positive utility conditional on that recommendation, following $a$ is strictly better than deviating to $\hat{a}$. If instead $\hat{a}\in F$, since $a\in\supp(x)$, the Nash equilibrium condition (\ref{eq: NE condition}) and the definition of $\tilde{u}(a,a')$ imply
\[
\sum_{a'\in H(a)}\pi(a,a')x[a'] \geq \sum_{a'\in H(\hat a)}\pi(\hat a,a')x[a'].
\]
Consequently,
\begin{align*}
\sum_{a'\in A}\tilde{\alpha}[a,a']
\left[u_\theta(a,a',\theta)-u_\theta(\hat a,a',\theta)\right] &=
\frac{x[a]}{P} \left[ \sum_{a'\in H(a)}\pi(a,a')x[a'] - \sum_{a'\in H(a)\cap H(\hat a)} \pi(\hat a,a')x[a'] \right]\\
&\geq \frac{x[a]}{P} \left[ \sum_{a'\in H(a)}\pi(a,a')x[a'] - \sum_{a'\in H(\hat a)} \pi(\hat a,a')x[a']\right]\\
&\geq 0,
\end{align*}
where the equality comes from the definitions of $\tilde{\alpha}$ and $u_\theta$, and the first inequality is due to the positivity of material payoffs. The above two cases together imply that for any function $g: A\rightarrow A$, we have
\[
\mathbb{E}_{\tilde{\alpha}} [u_{\theta}(a,a',\theta)]-\mathbb{E}_{\tilde{\alpha}} [u_{\theta}(g(a),a',\theta)]\geq 0.
\]
Since $\tilde{\alpha}$ is symmetric, the same conclusion holds for $\rho(\tilde{\alpha})$. Therefore, the incentive constraints for both sides are satisfied, meaning that $\tilde{\alpha}\in\CE_{\theta,\theta}$.
\end{proof}

Let $\theta$ denote a morally constrained selfish type. Lemma \ref{lemma: efficient CE} ensures that there exists a correlated equilibrium $\tilde{\alpha}\in \CE_{\theta,\theta}$ such that $\tilde{\alpha}[\E]=1$. Because the game between two type-$\theta$ agents is symmetric, the role-reversed agreement $\rho(\tilde{\alpha})$ also belongs to $\CE_{\theta,\theta}$. Moreover, the set of correlated equilibria is defined by linear incentive constraints and is therefore convex. Hence, we have 
\[\bar{\alpha}\equiv\frac{1}{2}\tilde{\alpha}+\frac{1}{2}\rho(\tilde{\alpha})\in \CE_{\theta,\theta}.\]
Since both $\tilde{\alpha}$ and $\rho(\tilde{\alpha})$ are efficient, so is $\bar{\alpha}$; moreover, the symmetrized $\bar{\alpha}$ gives both agents an expected utility of $m/2$.

Fix any competing type $\tau$ and any $\varepsilon\in(0,1)$. We claim that, in every complete-information stable outcome $(\mu,\gamma)$ under the population state $(\theta,\tau,\varepsilon)$, almost every type-$\theta$ agent obtains expected material payoff at least $m/2$. Suppose otherwise. Then a positive mass of type-$\theta$ agents obtain expected material payoff strictly below $m/2$. Since the utility of a morally constrained selfish agent is weakly below her material payoff, these agents also obtain expected utility strictly below $m/2$. Any two such agents can therefore rematch with one another and coordinate on $\bar{\alpha}$, which gives both agents expected utility $m/2$. This contradicts external stability. Hence, $G_\theta(\mu,\gamma)\geq m/2$.

Because each match generates at most $m$ in total material payoff and the unit-mass population contains a total mass of $1/2$ of matches, the aggregate material payoff of the population cannot exceed $m/2$. Hence,
\[
(1-\varepsilon)G_\theta(\mu,\gamma) +\varepsilon G_\tau(\mu,\gamma) \leq \frac{m}{2}.
\]
Combining this bound with the preceding result that $G_\theta(\mu,\gamma)\geq m/2$ yields
\[
G_\theta(\mu,\gamma)\geq\frac{m}{2}\geq G_\tau(\mu,\gamma).
\]
Thus, the morally constrained selfish type $\theta$ is evolutionarily dominant.

\subsubsection{Proof of Proposition \ref{prop: parochial efficient}}\label{proof: parochial efficient}
Let $\theta$ denote a parochial efficient type. First observe that any efficient agreement $\alpha$ constitutes a correlated equilibrium between two type-$\theta$ agents. Indeed, conditional on any recommendation, following it already yields the maximum joint material payoff, so no unilateral deviation can further increase utility. 


Fix any competing type $\tau$ and any $\varepsilon\in(0,1)$. We first show that every complete-information stable outcome $(\mu,\gamma)$ under the population state $(\theta,\tau,\varepsilon)$ must exhibit perfectly assortative matching, i.e., $\mu_{\theta}[\theta]=\mu_{\tau}[\tau]=1$. By contradiction, suppose instead that $\mu_{\theta}[\tau]>0$. Then a positive mass of type-$\theta$ agents obtain an expected utility of 0. Two such agents can rematch with one another and coordinate on an efficient agreement $\tilde{\alpha}$. Since the efficient agreement $\tilde{\alpha}$ is indeed a correlated equilibrium between two type-$\theta$ agents, which yields expected utility $\mathbb{E}_{\tilde{\alpha}}[\pi(a, a')+\pi(a', a)]=m>0$ to both sides, external stability is violated.

We next show that almost every match between two type-$\theta$ agents exhibits efficient play. Suppose some $\alpha\in\supp(\gamma_{\theta,\theta})$ is inefficient. Then the type-$\theta$ agents in these matches obtain expected utility strictly below $m$. Any two such agents can profitably rematch and coordinate on an efficient agreement, again contradicting external stability. Hence, every $\alpha\in\supp(\gamma_{\theta,\theta})$ satisfies $\alpha[\E]=1$. We then have
\begin{align*}
G_\theta(\mu,\gamma)
&=\int_{\A}\mathbb E_\alpha[\pi(a,a')]\mathop{d\gamma_{\theta,\theta}[\alpha]}\\
&=\int_{\A}\mathbb E_\alpha\left[\frac{\pi(a,a')+\pi(a',a)}{2}\right]\mathop{d\gamma_{\theta,\theta}[\alpha]}\\
&=\frac{m}{2}.
\end{align*}
The second equality follows from the exchangeability of $\gamma_{\theta,\theta}$, and the last from the efficiency of every agreement in its support.

Finally, perfect assortativity implies that type-$\tau$ agents are matched only among themselves. Since the joint material payoff cannot exceed $m$, we have $G_\tau(\mu,\gamma)\leq m/2$. Therefore,
\[
G_\theta(\mu,\gamma)=\frac{m}{2}\geq G_\tau(\mu,\gamma).
\]
Thus, the parochial efficient type $\theta$ is evolutionarily dominant.

\subsubsection{Proof of Proposition \ref{prop: necessary with ci}} \label{proof: necessary with ci}
Fix a preference type $\theta$, and suppose that
\[\hat{\alpha}\in\argmax_{\alpha\in \CE_{\theta,\theta}}\,\mathbb{E}_\alpha[u_{\theta}(a,a',\theta)+u_{\theta}(a',a,\theta)]\]
is inefficient. Let $\tau$ be a parochial efficient type, and fix any $\varepsilon\in(0,1)$. By Proposition \ref{prop: parochial efficient}, every complete-information stable outcome $(\mu,\gamma)$ under the population state $(\theta,\tau,\varepsilon)$ satisfies $G_\tau(\mu,\gamma)\geq G_\theta(\mu,\gamma)$.

It remains to show that the inequality is strict for some complete-information stable outcome. To this end, choose an outcome $(\mu,\gamma)$ as follows: let the matching profile be perfectly assortative, i.e., $\mu_\theta[\theta]=\mu_\tau[\tau]=1$; for type-$\theta$ matches, set $\gamma_{\theta,\theta}[\bar{\alpha}]=1$ where $\bar{\alpha}=\hat{\alpha}/2+\rho(\hat{\alpha})/2$; for type-$\tau$ matches, let all agents coordinate on some efficient agreement $\tilde{\alpha}$. This outcome is internally stable because the symmetry and convexity of $\CE_{\theta,\theta}$ imply that $\bar{\alpha}\in \CE_{\theta,\theta}$, while every efficient agreement is a correlated equilibrium between two parochial efficient agents.

We next verify external stability of $(\mu,\gamma)$. Type-$\tau$ agents already obtain their highest possible utility $m$, so they cannot participate in a blocking pair. No pair of type-$\theta$ agents can profitably rematch either. Suppose, to the contrary, that they can coordinate on some agreement ${\alpha'}\in\CE_{\theta,\theta}$ that makes both sides strictly better off. Since every type-$\theta$ agent's current utility is $\mathbb{E}_{\hat{\alpha}}[u_{\theta}(a,a',\theta)+u_{\theta}(a',a,\theta)]/2$, a profitable deviation would require
\[
\min\left\{\mathbb E_{{\alpha'}}[u_\theta(a,a',\theta)],\mathbb E_{\rho({\alpha'})}[u_\theta(a,a',\theta)]\right\}>\frac{\mathbb{E}_{\hat{\alpha}}[u_{\theta}(a,a',\theta)+u_{\theta}(a',a,\theta)]}{2}.
\]
But this implies
\[
\mathbb{E}_{{\alpha'}}[u_{\theta}(a,a',\theta)+u_{\theta}(a',a,\theta)]>\mathbb{E}_{\hat{\alpha}}[u_{\theta}(a,a',\theta)+u_{\theta}(a',a,\theta)],
\]
contradicting the choice of $\hat{\alpha}$. Hence, the constructed outcome is complete-information stable.

Finally, type-$\tau$ agents obtain average material payoff $m/2$. By contrast,
\[
G_\theta(\mu,\gamma) = \frac{1}{2}\mathbb E_{\bar{\alpha}}[\pi(a,a')+\pi(a',a)] < \frac{m}{2},
\]
where the strict inequality follows because $\bar{\alpha}=\hat{\alpha}/2+\rho(\hat{\alpha})/2$ and $\hat{\alpha}$ is inefficient. Therefore,
\[
G_\tau(\mu,\gamma)=\frac{m}{2}>G_\theta(\mu,\gamma).
\]
Thus, the preference type $\theta$ is evolutionarily dominated.

\subsubsection{Proof of Proposition \ref{prop: negative result}}\label{proof: negative result}
(i) Let $\theta$ denote a selfish type, and suppose there exists an efficient agreement $\tilde{\alpha}\in\CE_{\theta,\theta}$. By the symmetry and convexity of $\CE_{\theta,\theta}$, the agreement defined by $\bar{\alpha}\equiv \tilde{\alpha}/2+\rho(\tilde{\alpha})/2$ is also a correlated equilibrium between two selfish agents. Moreover, the symmetrized $\bar{\alpha}$ is efficient and gives each agent an expected material payoff of $m/2$.

Fix any competing type $\tau$, any $\varepsilon\in(0,1)$, and any complete-information stable outcome $(\mu,\gamma)$ under the population state $(\theta,\tau,\varepsilon)$. We claim that almost every type-$\theta$ agent obtains an expected material payoff of at least $m/2$. To see this, suppose otherwise. Then a positive mass of type-$\theta$ agents obtain expected material payoffs strictly below $m/2$. Since the utility of selfish agents coincides with their material payoff, these agents can block the outcome by rematching with one another and coordinating on $\bar{\alpha}$, a contradiction. Therefore, we have $G_\theta(\mu,\gamma)\geq m/2$ for all complete-information stable outcomes. 

Because the aggregate material payoff cannot exceed $m/2$, we have $(1-\varepsilon)G_\theta(\mu,\gamma) +\varepsilon G_\tau(\mu,\gamma) \leq m/2$. Together with $G_\theta(\mu,\gamma)\geq m/2$, this gives
\[
G_\theta(\mu,\gamma)\geq\frac{m}{2}\geq G_\tau(\mu,\gamma).
\]
Because $\tau$, $\varepsilon$, and $(\mu,\gamma)$ were arbitrary, the selfish type $\theta$ is evolutionarily dominant.

Conversely, suppose no agreement in $\CE_{\theta,\theta}$ is efficient. This implies that every agreement
\[\hat{\alpha}\in\argmax_{\alpha\in \CE_{\theta,\theta}}\,\mathbb{E}_\alpha[u_{\theta}(a,a',\theta)+u_{\theta}(a',a,\theta)]\]
is not efficient. Proposition \ref{prop: necessary with ci} then implies that $\theta$ is evolutionarily dominated.

(ii) Let $\theta$ denote an efficient type. Suppose first that every action pair in $\E$ gives the two agents equal material payoffs. Fix any competing type $\tau$, any $\varepsilon\in(0,1)$, and any complete-information stable outcome $(\mu,\gamma)$ under the population state $(\theta,\tau,\varepsilon)$.

We claim that almost every type-$\theta$ agent obtains expected utility $m$. To see this, suppose not. Then a positive mass of type-$\theta$ agents obtain expected utility strictly below $m$, and these agents can form a blocking pair with one another and coordinate on any efficient agreement, contradicting external stability. It follows that almost every type-$\theta$ agent interacts efficiently. Because every efficient action pair divides the material payoff equally, $G_\theta(\mu,\gamma)=m/2$. The aggregate payoff bound then implies
\[
G_\theta(\mu,\gamma)=\frac{m}{2}\geq G_\tau(\mu,\gamma).
\]
This means the efficient type $\theta$ is evolutionarily dominant.

For the converse, suppose that some efficient action pair gives the two agents unequal material payoffs. Choose
\[
(a^*,a^\dagger)\in
\argmax_{(a,a')\in\E}\pi(a,a').
\]
Because $\E$ is symmetric and contains an unequal payoff division, $\pi(a^*,a^\dagger)>m/2>\pi(a^\dagger,a^*)$. Let $\tau$ denote the morally constrained selfish type.

Fix any $\varepsilon\in(0,1)$ and consider an arbitrary complete-information stable outcome $(\mu,\gamma)$ under the population state $(\theta,\tau,\varepsilon)$. The argument establishing Proposition \ref{prop: morally constrained selfish} implies that type-$\tau$ agents obtain an average material payoff of at least $m/2$. The aggregate payoff bound therefore yields $G_\tau(\mu,\gamma)\geq m/2 \geq G_\theta(\mu,\gamma)$.

It remains to show that the inequality is strict in some stable outcome. Choose
\[
c\in(0,\min\{\varepsilon,1-\varepsilon\})
\]
and define the matching profile by $\mu_\theta[\tau]=c/{(1-\varepsilon)}$ and $\mu_\tau[\theta]=c/{\varepsilon}$, with the remaining probability assigned to same-type matches. The consistency condition holds as $(1-\varepsilon)\mu_\theta[\tau]=\varepsilon\mu_\tau[\theta]=c$. Let $\bar{\alpha}\in\CE_{\tau,\tau}$ be a symmetric efficient agreement, whose existence follows from the argument establishing Proposition \ref{prop: morally constrained selfish}. Because $\bar{\alpha}$ is efficient, it also belongs to $\CE_{\theta,\theta}$. Let $\gamma_{\theta,\theta}[\bar{\alpha}]=\gamma_{\tau,\tau}[\bar{\alpha}]=1$. For cross-type matches, $\gamma_{\theta,\tau}[\hat{\alpha}]=1$, where $\hat{\alpha}[a^\dagger,a^*]=1$.

The outcome is internally stable. In every cross-type match, the type-$\theta$ agent obtains utility $m$, her highest feasible utility. The type-$\tau$ agent also finds it optimal to follow the agreement: any deviation that results in inefficient play gives her utility zero, whereas any deviation that preserves efficiency gives her a material payoff no greater than $\pi(a^*,a^\dagger)$, by the choice of $(a^*,a^\dagger)$.

To verify external stability, first observe that every type-$\theta$ agent obtains utility $m$ and therefore cannot strictly benefit from any deviation. Hence, no blocking pair can involve a type-$\theta$ agent. Every type-$\tau$ agent obtains utility at least $m/2$: she obtains exactly $m/2$ in a same-type match and $\pi(a^*,a^\dagger)>m/2$ in a cross-type match. Moreover, under any agreement $\alpha$ between two type-$\tau$ agents,
\[
\mathbb{E}_{\alpha}[u_\tau(a,a',\tau)]
+\mathbb{E}_{\rho(\alpha)}[u_\tau(a,a',\tau)]
=m\alpha[\E]\leq m.
\]
Thus, no agreement can make two type-$\tau$ agents, each of whom already obtains at least $m/2$, strictly better off. No blocking pair exists, so the constructed outcome is complete-information stable.

Because cross-type matches occur with mass $c>0$ and $\pi(a^*,a^\dagger)>m/2>\pi(a^\dagger,a^*)$,
we have
\[
G_\tau(\mu,\gamma)>\frac{m}{2}>G_\theta(\mu,\gamma).
\]
Since $\varepsilon$ was arbitrary, the efficient type $\theta$ is evolutionarily dominated.

\subsection{Proofs for Section \ref{sec: preference evolution incomplete}: Incomplete Information}

In this section, we first elaborate on Example \ref{eg: morally constrained dominated}. Next, we establish the positive result under incomplete information, Propositions \ref{prop: parochial efficient with ii}. Finally, we prove the negative result for the morally constrained selfish type, Proposition \ref{prop: morally constrained dominated}. We adopt this order because the proof of the positive result provides the central economic insights and develops a blocking argument that is also used to establish the negative result.

\subsubsection{More on Example \ref{eg: morally constrained dominated}}\label{proof of eg: morally constrained dominated}
We verify that the outcome $(\Lambda,p,q,\mu,\gamma)$ constructed in Example \ref{eg: morally constrained dominated} is incomplete-information stable. Internal stability is immediate, so we focus on external stability. Note that a necessary condition for blocking is that both targeted types strictly benefit from the deviation when no nontargeted types participate.

First, consider a pair consisting of one type-$\theta$ agent and one type-$\tau$ agent. Against a type-$\theta$ partner, $x$ is strictly dominant for the type-$\tau$ agent. Any viable deviating agreement must therefore prescribe $x$ to that agent, against which the type-$\theta$ agent can obtain at most $2$, her status quo utility. Hence, no such pair can block.

Second, consider two type-$\tau$ agents carrying either label. Every type-$\tau$ agent obtains utility of at least $6$ in the status quo. When only the targeted agents participate, both value the joint material payoff, which cannot exceed $6$. Hence, no pair of type-$\tau$ agents can block.

Finally, consider two type-$\theta_\lambda$ agents who target one another and propose an agreement $\hat{\alpha}$. Consider the deviation plan $(\lambda,D,d)$ satisfying $D=\{\theta,\tau\}$, $d(\theta)=f$, and $d(\tau)=g$, where $g(\cdot)=x$.\footnote{\label{ft: rational play}Such play is rational and strictly profitable for a type-$\tau_\lambda$ agent facing a type-$\theta$ partner: it yields utility $8$, rather than her status quo utility of $6$.} Given this plan for her deviating partner, consider a type-$\theta_\lambda$ agent who follows $\hat{\alpha}$. Conditional on being recommended $x$, she obtains expected utility at most $4q_\lambda[\theta]<2$. Conditional on being recommended $y$, the agent obtains utility at most $2$, regardless of her partner's type. Her expected utility from following $\hat{\alpha}$ therefore cannot exceed $2$, her status quo utility. Hence, no pair of type-$\theta_\lambda$ agents can block.

Therefore, no incomplete-information blocking pair exists, and the outcome is incomplete-information stable.

\subsubsection{Proof of Proposition \ref{prop: parochial efficient with ii}}\label{proof: parochial efficient with ii}
Let $\theta$ denote a parochial efficient type, and fix an arbitrary preference type $\tau\in\Theta$. Consider any incomplete-information stable outcome $(\Lambda,p,q,\mu,\gamma)$. The proof proceeds through two lemmas.

\begin{lemma}\label{lemma: perfect assortativity}
    In an incomplete-information stable outcome $(\Lambda,p,q,\mu,\gamma)$, if $\lambda\in\Lambda$ and $q_\lambda[\theta]>0$, then $\mu_\lambda[\lambda]=1$.
\end{lemma}
\begin{proof}
Suppose, toward a contradiction, that $\mu_\lambda[\lambda']>0$ for some $\lambda'\neq\lambda$. By the consistency condition, $\mu_{\lambda'}[\lambda]>0$ as well. Moreover, because distinct labels induce distinct type distributions, $q_\lambda[\theta]\neq q_{\lambda'}[\theta]$. Interchanging $\lambda$ and $\lambda'$ if necessary, assume that $q_\lambda[\theta]>q_{\lambda'}[\theta]$. Take any $\alpha'\in\supp(\gamma_{\lambda,\lambda'})$. We show that two type-$\theta_\lambda$ agents with current position $(\lambda',\alpha')$ can form an incomplete-information blocking pair by targeting one another and proposing an efficient agreement $\hat{\alpha}$.

Fix any deviation plan $(\lambda,D,d)$ for one agent's prospective partner such that $\theta\in D$ and $d(\theta)=f$. Because $u_\theta(\cdot,\cdot,\tau)=0$, for any deviating strategy $g: A\rightarrow A$,
\begin{align*}
 \mathbb{E}_{ \hat{\alpha}}[\mathbb{E}_{q_{\lambda}} [u_\theta(g(a), d(\cdot)(a'), \cdot)\,|\,D]]&=\begin{cases}
     \mathbb{E}_{\hat{\alpha}}[u_\theta(g(a), a', \theta)] q_{\lambda}[\theta] &\text{ if }\tau\in D,\\
     \mathbb{E}_{\hat{\alpha}}[u_\theta(g(a), a', \theta)] &\text{ if }\tau\notin D.
 \end{cases}
\end{align*}
Since $\hat{\alpha}$ is efficient, following its recommendation yields utility $m$ against a type-$\theta$ partner, the highest feasible utility. Hence, we have $f \in \argmax_{g}\,\mathbb{E}_{\hat{\alpha}}[u_\theta(g(a), a', \theta)]$, which in turn implies
\[
f\in \argmax_{g} \,\mathbb{E}_{ \hat{\alpha}}[\mathbb{E}_{q_{\lambda}} [u_\theta(g(a), d(\cdot)(a'), \cdot)\,|\,D]]
\]
Moreover, the expected utility from following $\hat{\alpha}$ in the deviation satisfies
\begin{align*}
    \mathbb{E}_{\hat{\alpha}}[\mathbb{E}_{q_{\lambda}} [u_\theta(a, d(\cdot)(a'), \cdot)\,|\,D]] & \geq \mathbb{E}_{\hat{\alpha}}[u_\theta(a, a', \theta)] q_{\lambda}[\theta]\\
    &= m q_{\lambda}[\theta]\\
    &>\mathbb{E}_{\alpha'}[u_\theta(a, a', \theta)] q_{\lambda'}[\theta] \\
    &=\mathbb{E}_{\alpha'}[\mathbb{E}_{q_{\lambda'}} [u_\theta(a, a', \cdot)]].
\end{align*}
The weak inequality above follows because $q_{\lambda}[\theta]\leq 1$ and $u_\theta(\cdot,\cdot,\theta)\geq0$; the strict inequality follows from $q_{\lambda}[\theta]>q_{\lambda'}[\theta]$ and $m>0$ being the maximum joint material payoff.

The same argument applies to the other targeted agent, who follows $\rho(\hat{\alpha})$. Thus, the two type-$\theta_\lambda$ agents form an incomplete-information blocking pair, contradicting the stability of $(\Lambda,p,q,\mu,\gamma)$. Therefore, $\mu_\lambda[\lambda']=0$ for every $\lambda'\neq\lambda$, meaning that $\mu_\lambda[\lambda]=1$.
\end{proof}

\begin{lemma}\label{lemma: efficient play}
    In an incomplete-information stable outcome $(\Lambda,p,q,\mu,\gamma)$, if $\lambda\in\Lambda$ and $q_\lambda[\theta]>0$, then any agreement $\alpha\in\supp(\gamma_{\lambda,\lambda})$ is efficient.
\end{lemma}
\begin{proof}
By contradiction, suppose that some $\alpha'\in\supp(\gamma_{\lambda,\lambda})$ is inefficient. We show that two type-$\theta_\lambda$ agents with current position $(\lambda,\alpha')$ can form an incomplete-information blocking pair by targeting one another and proposing an efficient agreement $\hat{\alpha}$.

The argument in the proof of Lemma \ref{lemma: perfect assortativity} implies that, under every deviation plan $(\lambda,D,d)$ for the partner such that $\theta\in D$ and $d(\theta)=f$, following $\hat{\alpha}$ is optimal for the deviating type-$\theta_\lambda$ agent. Moreover, the expected utility from following $\hat{\alpha}$ in the deviation satisfies
\begin{align*}
    \mathbb{E}_{\hat{\alpha}}[\mathbb{E}_{q_{\lambda}} [u_\theta(a, d(\cdot)(a'), \cdot)\,|\,D]] & \geq \mathbb{E}_{\hat{\alpha}}[u_\theta(a, a', \theta)] q_{\lambda}[\theta]\\
    &= m q_{\lambda}[\theta]\\
    &>\mathbb{E}_{\alpha'}[u_\theta(a, a', \theta)] q_{\lambda}[\theta] \\
    &=\mathbb{E}_{\alpha'}[\mathbb{E}_{q_{\lambda}} [u_\theta(a, a', \cdot)]],
\end{align*}
where the strict inequality now follows from $q_{\lambda}[\theta]>0$ and the fact that $\alpha'$ is inefficient.

The same argument applies to the other targeted agent, who follows $\rho(\hat{\alpha})$. Thus, the two type-$\theta_\lambda$ agents form an incomplete-information blocking pair, contradicting external stability. Hence, every agreement $\alpha\in\supp(\gamma_{\lambda,\lambda})$ must be efficient.
\end{proof}

The previous two lemmas imply that
\begin{align*}
    G_\theta(\Lambda, p, q, \mu,\gamma)&=\frac{1}{1-\varepsilon} \sum_{\lambda\in\Lambda}\left\{\int_{\A}\mathbb{E}_{\alpha}[\pi(a,a')]\mathop{d \gamma_{\lambda,\lambda}[\alpha]}\right\} p[\lambda]q_{\lambda}[\theta]\\
    &=\frac{1}{1-\varepsilon} \sum_{\lambda\in\Lambda}\left\{\int_{\A}\mathbb{E}_{\alpha}\left[\frac{\pi(a,a')+\pi(a',a)}{2}\right]\mathop{d \gamma_{\lambda,\lambda}[\alpha]}\right\}p[\lambda]q_{\lambda}[\theta]\\
    &=\frac{1}{1-\varepsilon}\sum_{\lambda\in\Lambda} \frac{m}{2} p[\lambda]q_{\lambda}[\theta]\\
    &=\frac{m}{2}.
\end{align*}
The first equality uses perfect assortativity, and the second comes from the exchangeability of $\gamma_{\lambda,\lambda}$. The third equality follows because, for every label $\lambda$ with $q_\lambda[\theta]>0$, every agreement in $\supp(\gamma_{\lambda,\lambda})$ is efficient. The final equality follows from the marginal condition.

Note that the aggregate material payoff of the population cannot exceed $m/2$, that is,
\[
    (1-\varepsilon)G_\theta(\Lambda, p, q, \mu,\gamma)+\varepsilon G_\tau(\Lambda, p, q, \mu,\gamma)\leq\frac{m}{2}.
\]
Thus, we must have 
\[
    G_\theta(\Lambda, p, q, \mu,\gamma)=\frac{m}{2}\geq G_\tau(\Lambda, p, q, \mu,\gamma).
\]
Since the competing type $\tau$, its population share $\varepsilon$, and the incomplete-information stable outcome were arbitrary, we can conclude that the parochial efficient type $\theta$ is evolutionarily dominant under incomplete information.

\subsubsection{Proof of Proposition \ref{prop: morally constrained dominated}}\label{proof: morally constrained dominated}
Let $\theta$ denote a morally constrained selfish type. Relabeling $a^*$ and $a^\dagger$ if necessary, assume that $\pi(a^*,a^\dagger)>\pi(a^\dagger,a^*)$, and consider a preference type $\tau$ with utility function
\begin{equation*}
   u_\tau(a,a',t) = \begin{cases}
    \mathbf{1}_{\{(a,a')\in\E\}} &\text{if $t=\tau$,}\\
    2\cdot \mathbf{1}_{\{a=a^*\}}  &\text{if $t\neq \tau$.}
    \end{cases}
\end{equation*}
Fix an arbitrary $\varepsilon\in(0,1)$ and consider the population state $(\theta,\tau,\varepsilon)$. We show that $\theta$ is evolutionarily dominated by $\tau$ through two lemmas. Because the blocking arguments parallel those used in the proof of Proposition \ref{prop: parochial efficient with ii}, we omit some details to avoid repetition.

\begin{lemma}\label{lemma: prop 5 weak}
    $G_\tau(\Lambda, p, q, \mu, \gamma)\geq G_\theta(\Lambda, p, q, \mu, \gamma)$ for all incomplete-information stable outcomes $(\Lambda, p, q, \mu, \gamma)$.
\end{lemma}
\begin{proof}
Fix an arbitrary incomplete-information stable outcome $(\Lambda, p, q, \mu, \gamma)$. We first record the following observation: if $q_\lambda[\theta]\leq 1/3$, almost every type-$\tau_\lambda$ agent must obtain expected utility of at least $1$. Otherwise, two such agents can target one another and propose an efficient agreement $\hat{\alpha}$ satisfying $\hat{\alpha}[a^*,a^{\dagger}]=\hat{\alpha}[a^{\dagger},a^*]=1/2$. Following $\hat{\alpha}$ yields utility $1$ whether or not the nontargeted type-$\theta_\lambda$ agents participate. If they participate, the only potentially profitable deviation is to choose $a^*$ when $a^\dagger$ is recommended. Such a deviation is unprofitable because $2q_\lambda[\theta]\leq (1-q_\lambda[\theta])$ as $q_\lambda[\theta]\leq 1/3$. The proposed deviation therefore constitutes an incomplete-information blocking pair, a contradiction.

Consider any $\lambda,\lambda'\in\Lambda$ such that $q_\lambda[\tau],q_{\lambda'}[\tau]>0$ and $\mu_{\lambda}[\lambda']>0$. We claim that for every $\alpha\in\supp(\gamma_{\lambda,\lambda'})$: (i) $\alpha[\E]=1$; and (ii) if $q_\lambda[\tau]>q_{\lambda'}[\tau]$, then $\alpha[a^*,a^{\dagger}]\geq\alpha[a^{\dagger},a^*]$. 

For (i), suppose there exists $\alpha\in \supp(\gamma_{\lambda,\lambda'})$ such that $\alpha[\E]<1$. We first show that $q_\lambda[\theta],q_{\lambda'}[\theta]\leq 1/3$. If $q_\lambda[\theta]> 1/3$, choosing $a^*$ gives a type-$\tau_{\lambda'}$ agent utility at least $2q_\lambda[\theta]$, whereas any other action gives utility at most $1-q_\lambda[\theta]$. Internal stability therefore requires the agreement to recommend $a^*$ to label-$\lambda'$ agents with probability one. But $a^\dagger$ is the only action that can be optimal against $a^*$ for type-$\theta_{\lambda}$ agents. Internal stability again implies $\alpha[a^{\dagger},a^*]=1$, a contradiction. Hence, $q_\lambda[\theta]\leq 1/3$. A symmetric argument gives $q_{\lambda'}[\theta]\leq 1/3$.

If $\alpha[\E]<1$, at least one of $\alpha[a^*,a^{\dagger}]$ and $\alpha[a^{\dagger},a^*]$ is strictly below $1/2$. Suppose $\alpha[a^*,a^{\dagger}]<1/2$ without loss. We consider two cases. If $q_{\lambda'}[\theta]=0$, then $\mathbb{E}_\alpha[\mathbb{E}_{q_{\lambda'}}[u_{\tau}(a,a',\cdot)]]=\alpha[\E]<1$. If $q_{\lambda'}[\theta]>0$, the outcome $(a^*,a^\dagger)$ yields utility $1+q_{\lambda'}[\theta]$, while every other action pair yields at most $1-q_{\lambda'}[\theta]$. In particular, an inefficient pair in which the agent plays $a^*$ yields $2q_{\lambda'}[\theta]\leq 1-q_{\lambda'}[\theta]$ as $q_{\lambda'}[\theta]\leq 1/3$. Therefore,
\begin{align*}
\mathbb{E}_\alpha[\mathbb{E}_{q_{\lambda'}}[u_\tau(a,a',\cdot)]]&\leq\alpha[a^*,a^\dagger](1+q_{\lambda'}[\theta])+(1-\alpha[a^*,a^\dagger])(1-q_{\lambda'}[\theta])\\
&= 1+(2\alpha[a^*,a^\dagger]-1)q_{\lambda'}[\theta]\\
&<1,
\end{align*}
where the strict inequality comes from $\alpha[a^*,a^{\dagger}]<1/2$ and $q_{\lambda'}[\theta]>0$. Because expected utility is continuous in the agreement, this strict inequality holds in a neighborhood of $\alpha$. Since $\alpha\in\supp(\gamma_{\lambda,\lambda'})$, a positive mass of type-$\tau_\lambda$ agents obtain expected utility strictly below $1$ in either case. This, together with $q_\lambda[\theta]\leq 1/3$, contradicts the preliminary observation. This proves (i).

For (ii), suppose that $q_\lambda[\tau]>q_{\lambda'}[\tau]$ and $\alpha[a^*,a^{\dagger}]<\alpha[a^{\dagger},a^*]$. Because (i) holds and expected utility is continuous in the agreement, a positive mass of type-$\tau_{\lambda}$ agents obtain expected utility $q_{\lambda'}[\tau]+2\alpha[a^*,a^{\dagger}](1-q_{\lambda'}[\tau])<1$. Because \(\alpha[a^\dagger,a^*]>0\), obedience when $a^\dagger$ is recommended requires $q_{\lambda'}[\theta]\leq1/3$. Moreover, $q_\lambda[\tau]>q_{\lambda'}[\tau]$ implies $q_\lambda[\theta]<q_{\lambda'}[\theta]$. Hence, $q_\lambda[\theta]\leq 1/3$, contradicting the preliminary observation. This proves (ii).

Next, we show that any label containing only type $\theta$ must be matched with itself. By contradiction, suppose $q_{\lambda_{\theta}}[\theta]=1$ and $\mu_{\lambda_{\theta}}[\lambda]>0$ for some $\lambda\neq\lambda_{\theta}$. Because distinct labels induce distinct type distributions, $q_{\lambda}[\tau]>0$. A type-$\tau_\lambda$ agent facing a type-$\theta_{\lambda_\theta}$ partner must play $a^*$, so the latter's current utility is at most $\pi(a^\dagger,a^*)<m/2$. Two type-$\theta_{\lambda_\theta}$ agents can instead target one another and play $\hat{\alpha}$, which gives each of them utility $m/2$. This forms an incomplete-information blocking pair. Therefore, $\mu_{\lambda_\theta}[\lambda_\theta]=1$.

It remains to compare average material payoffs. Define
\[
    \Pi_{\lambda,\lambda'}=\int_{\A}\mathbb{E}_{\alpha}[\pi(a,a')]\mathop{d \gamma_{\lambda,\lambda'}[\alpha]}.
\]
Consider any $\lambda,\lambda'\in\Lambda$ such that $\mu_\lambda[\lambda']>0$ and $q_\lambda[\tau]+q_{\lambda'}[\tau]>0$. Because every pure-$\theta$ label is matched with itself, we must have $q_\lambda[\tau],q_{\lambda'}[\tau]>0$. Claim (i) therefore implies $\Pi_{\lambda,\lambda'}+\Pi_{\lambda',\lambda}=m$. Moreover, Claim (ii) implies that the label with a higher proportion of type-$\tau$ agents obtains a weakly higher material payoff. Hence,
\[
\Pi_{\lambda,\lambda'}q_\lambda[\tau] +\Pi_{\lambda',\lambda}q_{\lambda'}[\tau] \geq (\Pi_{\lambda,\lambda'} +\Pi_{\lambda',\lambda})\frac{q_\lambda[\tau]+q_{\lambda'}[\tau]}{2}=(q_\lambda[\tau]+q_{\lambda'}[\tau])\frac{m}{2}.
\]
Therefore,
\begin{align*}
G_\tau(\Lambda,p,q,\mu,\gamma)
&= \frac{1}{2\varepsilon}\sum_{\lambda,\lambda'\in\Lambda} p[\lambda]\mu_{\lambda}[\lambda'] (\Pi_{\lambda,\lambda'}q_\lambda[\tau] + \Pi_{\lambda',\lambda}q_{\lambda'}[\tau])\\
&\geq \frac{1}{2\varepsilon}\sum_{\lambda,\lambda'\in\Lambda} p[\lambda]\mu_{\lambda}[\lambda'](q_\lambda[\tau]+q_{\lambda'}[\tau])\frac{m}{2}\\
&=\frac{1}{\varepsilon}\sum_{\lambda\in\Lambda}p[\lambda]q_\lambda[\tau]\frac{m}{2}\\
&=\frac{m}{2}.
\end{align*}
The first equality uses the consistency condition, while the penultimate equality additionally uses the fact that $\sum_{\lambda'\in\Lambda}\mu_\lambda[\lambda']=1$. The final equality follows from the marginal condition. Because the aggregate material payoff of the population cannot exceed $m/2$, it follows that
\[
G_\theta(\Lambda,p,q,\mu,\gamma) \leq \frac{m}{2} \leq G_\tau(\Lambda,p,q,\mu,\gamma),
\]
as desired.
\end{proof}

\begin{lemma}\label{lemma: prop 5 strict}
    $G_\tau(\Lambda, p, q, \mu, \gamma)> G_\theta(\Lambda, p, q, \mu, \gamma)$ for some incomplete-information stable outcome $(\Lambda, p, q, \mu, \gamma)$.
\end{lemma}
\begin{proof}

Fix the population state $(\theta,\tau,\varepsilon)$ and consider an outcome $(\Lambda, p, q, \mu,\gamma)$ as follows. Choose $q_\lambda[\theta]$ such that
\[0<q_\lambda[\theta]\leq \frac{\pi(a^\dagger,a^*)}{\pi(a^*,a^\dagger)} \quad\text{and}\quad q_\lambda[\theta]<2(1-\varepsilon).\]
There are three labels $\Lambda=\{\lambda_\theta,\lambda,\lambda_\tau\}$, where $q_{\lambda_\theta}[\theta]=q_{\lambda_\tau}[\tau]=1$. Set $p[\lambda]=p[\lambda_{\tau}]=\varepsilon/(2-q_{\lambda}[\theta])$, and let $\mu_{\lambda_\theta}[\lambda_\theta]=\mu_{\lambda}[\lambda_\tau]=\mu_{\lambda_\tau}[\lambda]=1$. Finally, let $\gamma_{\lambda_\theta,\lambda_\theta}[\hat{\alpha}]=1$ where $\hat{\alpha}[a^*,a^\dagger]=\hat{\alpha}[a^\dagger,a^*]=1/2$ and $\gamma_{\lambda,\lambda_\tau}[\tilde{\alpha}]=1$ where $\tilde{\alpha}[a^\dagger,a^*]=1$. Figure \ref{proof pic: homophilic efficient unstable} illustrates such an outcome.

\begin{figure}[!ht]
    \centering
    \begin{tikzpicture}[scale=0.85]
    \draw[black!80] (-4,-0.4) rectangle +(8,0.8);
    
    \fill[red, fill opacity=0.2] (1.8,-0.4) rectangle +(2.2,0.8);
    \node at (3.4,0) {$\lambda_\tau$};
    
    \fill[blue, fill opacity=0.2] (-4,-0.4) rectangle +(5.8,0.8);
    \node at (-1.1,0) {$\lambda_\theta$};
    
    \node at (2.2,0) {$\lambda$};

    \draw [decorate,decoration={brace,amplitude=5pt,mirror},xshift=0pt,yshift=-8pt]
    (1.8,-0.4) -- (4,-0.4) node [midway,yshift=-12pt, xshift=0pt] 
    {$\tau$};
    
    \draw [decorate,decoration={brace,amplitude=5pt,mirror},xshift=0pt,yshift=-8pt]
    (-4,-0.4) -- (1.8,-0.4) node [midway,yshift=-12pt, xshift=0pt] 
    {$\theta$};
    
    \draw[black!40, rounded corners, dashed] (-4.1,-0.5) rectangle (1.6,0.5);
    \fill[pattern=dots, pattern color=black!30,rounded corners] (-4.1,-0.5) rectangle (1.6,0.5);

    \draw[black!40, rounded corners, dashed] (1.6,-0.5) rectangle (2.8,0.5);
    \fill[pattern=north east lines, pattern color=black!30,rounded corners] (1.6,-0.5) rectangle (2.8,0.5);

    \draw[black!40, rounded corners, dashed] (2.8,-0.5) rectangle (4.1,0.5);
    \fill[pattern=dots, pattern color=black!30,rounded corners] (2.8,-0.5) rectangle (4.1,0.5);
    
    \draw [<-] (3.4,0.6) to [out=100, in=0] (2.8,1.4);
    \draw [->] (2.8,1.4) to [out=180, in=80] (2.2,0.6);
    \node at (2.8,1.4) [above=] {$\tilde{\alpha}$};
    
    \draw [<-] (-0.9,0.6) to [out=30, in=270] (-0.7,1);
    \draw [-] (-0.7,1) to [out=90, in=0] (-1.1,1.4);
    \draw [-] (-1.1,1.4) to [out=180, in=90] (-1.5,1);
    \draw [->] (-1.5,1) to [out=270, in=150] (-1.3,0.6);
    \node at (-1.1,1.4) [above=] {$\hat{\alpha}$};

    \end{tikzpicture}
    \caption{The outcome constructed in the proof of Lemma \ref{lemma: prop 5 strict}.}
    \label{proof pic: homophilic efficient unstable}
\end{figure}

We argue that this outcome is incomplete-information stable. Internal stability is straightforward, so we focus on external stability. No type-$\theta$ agent can form a blocking pair with a type-$\tau$ agent. When only the targeted agents participate, the type-$\tau$ agent strictly prefers $a^*$, to which $a^\dagger$ is the unique optimal response for the type-$\theta$ agent. The latter therefore obtains $\pi(a^\dagger,a^*)$, which is no greater than her status quo utility. Similarly, two type-$\tau$ agents cannot form a blocking pair because their utility from interacting with one another cannot exceed $1$.

It remains to consider pairs consisting of two type-$\theta$ agents. Two type-$\theta_{\lambda_\theta}$ agents cannot block because each already obtains $m/2$, while their joint utility cannot exceed $m$. Now consider any pair involving a targeted type-$\theta_\lambda$ agent, and focus on the targeted agent whose prospective partner has label $\lambda$. For any proposed agreement $\alpha'$, consider the deviation plan $(\lambda,D,d)$ satisfying $D=\{\theta,\tau\}$, $d(\theta)=f$, and $d(\tau)=g$, where $g(\cdot)=a^*$.\footnote{\label{ft: rational play 2}Such play is rational and strictly profitable for a type-$\tau_\lambda$ agent paired with a type-$\theta$ agent: it yields utility $2$, compared with her status quo utility of $1$.} Given this plan for her deviating partner, the agent's expected utility from following $\alpha'$ can be bounded recommendation by recommendation. Conditional on being recommended $a^*$, she obtains at most
\[
q_\lambda[\theta]\pi(a^*,a^\dagger) \leq \pi(a^\dagger,a^*).
\]
Conditional on being recommended $a^\dagger$, the agent obtains at most $\pi(a^\dagger,a^*)$, regardless of the partner's type. Any other recommendation yields zero utility. It follows that her expected utility from following $\alpha'$ cannot exceed $\pi(a^\dagger,a^*)$. The resulting utility is no greater than the status quo utility of a type-$\theta_\lambda$ agent and is strictly below that of a type-$\theta_{\lambda_\theta}$ agent. Thus, no blocking pair can involve a type-$\theta_\lambda$ partner.

Hence, the outcome $(\Lambda, p, q, \mu, \gamma)$ is incomplete-information stable. It is left to verify that $G_\tau(\Lambda, p, q, \mu, \gamma) > m/2 > G_\theta(\Lambda, p, q, \mu, \gamma)$. To see this, note first that play is efficient across all matches. Second, more than half of type-$\tau$ agents occupy the advantageous side of efficient play, choosing $a^*$ against $a^\dagger$, whereas more than half of type-$\theta$ agents occupy the disadvantaged side.
\end{proof}

\subsection{Proofs for Section \ref{sec: polymorphism}: Polymorphism}

\subsubsection{Proof of Proposition \ref{prop: existence}}
\label{proof: existence}

We reformulate our model as a roommate market in the sense of \citet{carmona2024}. We then show how a stable roommate matching in their framework induces a complete-information stable outcome in ours. Finally, we state and apply their existence result.

Given the primitives of our model, define a \textbf{roommate market} $\mathcal{E}=(\Theta_\nu, \nu, C, \mathbb{C}, (\succ_t)_{t\in\Theta_\nu})$ as follows. Fix an arbitrary strict total order $>$ on $\Theta_\nu$. For each $t\in\Theta_\nu$, define the set of \textbf{fair correlated equilibria} between two type-$t$ agents by\footnote{The formal definition of a roommate market in \citet{carmona2024} requires preferences under a given contract to be invariant to the two roles. We therefore restrict contracts between same-type agents to fair correlated equilibria. This restriction is without loss for stable outcomes: if an agreement gave the two roles different expected utilities, two agents occupying the less favorable role could rematch under its symmetrization and both strictly gain.}
\[
    \CE^F_{t,t}=\left\{\alpha\in\CE_{t,t}: \mathbb{E}_{\alpha}[u_t(a,a',t)] = \mathbb{E}_{\rho(\alpha)}[u_t(a,a',t)] \right\}.
\]
This set is nonempty. Indeed, for any $\alpha\in\CE_{t,t}$, convexity of $\CE_{t,t}$ and its invariance under role reversal imply that $\bar{\alpha}= \alpha/2 +\rho(\alpha)/2\in\CE^F_{t,t}$.

Let $\emptyset$ denote an unmatched partner and write $\bar{\Theta}_\nu=\Theta_\nu\cup\{\emptyset\}$. The \textbf{contract space} is $C=\A$. Define the \textbf{contract correspondence} $\mathbb{C}:\Theta_\nu\times\bar{\Theta}_\nu\rightrightarrows C$ by
\[
    \mathbb{C}(t,t')
    =
    \begin{cases}
        \CE_{t,t'} & \text{if $t,t'\in\Theta_\nu$ and $t>t'$,}\\
        \CE_{t',t} & \text{if $t,t'\in\Theta_\nu$ and $t<t'$,}\\
        \CE^F_{t,t} & \text{if $t=t'$,}\\
        \A & \text{if $t'=\emptyset$.}
    \end{cases}
\]
In particular, we have $\mathbb{C}(t,t') = \mathbb{C}(t',t)$ for all $t,t'\in\Theta_\nu$. Choose $\underline{u} < \min_{a,a',t,t'} u_t(a,a',t')$. For each type $t\in\Theta_\nu$, let $\succ_t$ be a binary relation on $\widebar{\Theta}_\nu\times C$ induced by the following utility function:
\[
    v_t(t',\alpha)=
    \begin{cases}
        \mathbb{E}_{\alpha}[u_t(a,a',t')] & \text{if $t\geq t'$,}\\
        \mathbb{E}_{\rho(\alpha)}[u_t(a,a',t')] & \text{if $t<t'$,}\\
        \underline{u} & \text{if $t'=\emptyset$.}
    \end{cases}
\]
By construction, these preferences only depend on the opponent's type and contract, but not on the role of the agents.

A \textbf{roommate matching} is a finite Borel measure $\varphi\in \mathcal{M}(\Theta_\nu\times \widebar{\Theta}_\nu\times C)$ such that $(t,t',\alpha)\in\supp(\varphi)$ implies $\alpha\in\mathbb{C}(t,t')$ and, for every $t\in\Theta_\nu$,
\begin{equation}
\label{eq: roommate accounting}
    \nu[t]=\varphi[\{t\}\times\Theta_\nu\times\A] + \varphi [\Theta_\nu\times\{t\}\times\A] + \varphi [\{t\}\times\{\emptyset\}\times\A].
\end{equation}
The three terms are the masses of type-$t$ agents recorded on the first side, on the second side, and as unmatched, respectively.

For each $t\in\Theta_\nu$, define the set of type $t$'s current positions by
\begin{align*}
    P_t(\varphi) = &\left\{ (t',\alpha)\in\Theta_\nu\times\mathcal{A}: (t,t',\alpha)\in\supp(\varphi) \text{ or }  (t',t,\alpha)\in\supp(\varphi) \right\}\\
    &\cup \left\{ (\emptyset,\alpha): (t,\emptyset,\alpha)\in\supp(\varphi)\right\}.
\end{align*}
The targets of a type-$t$ agent are
\[
    T_t(\varphi)=\{(t',\alpha)\in\Theta_\nu\times\mathcal{A}:
    \alpha\in\mathbb{C}(t,t') \text{ and } \exists p'\in P_{t'}(\varphi) \text{ s.t. }(t,\alpha)\succ_{t'}p' \}.
\]
Thus, $(t',\alpha)\in T_t(\varphi)$ if a type-$t$ agent can attract a type-$t'$ partner under contract $\alpha$. Let $\bar{T}_t(\varphi)=T_t(\varphi)\cup(\{\emptyset\}\times\mathcal{A})$. A roommate matching $\varphi$ is \textbf{stable} if there do not exist
$t\in\Theta_\nu$, $p\in P_t(\varphi)$, and
$\hat{p}\in\bar{T}_t(\varphi)$ such that $\hat{p}\succ_t p$.

The stability notions in the two frameworks are equivalent. For our existence result, however, only the following direction is needed.

\begin{lemma}\label{lemma: equivalent stability}
Suppose there exists a stable roommate matching in the roommate market $\mathcal{E}$. Then there exists a complete-information stable outcome under population distribution $\nu$.
\end{lemma}

\begin{proof}
Let $\varphi$ be a stable roommate matching. No unmatched position can belong to its support. Otherwise, suppose $(\emptyset,\alpha)\in P_t(\varphi)$ and choose any $\bar{\alpha}\in\CE^F_{t,t}$. Because both roles under $\bar{\alpha}$ yield utility strictly above $\underline{u}$, two unmatched type-$t$ agents could match with one another under $\bar{\alpha}$, contradicting the stability of $\varphi$.

Next, we construct an outcome $(\mu,\gamma)$ in our setting from the roommate matching $\varphi$. For $t,t'\in\Theta_\nu$, define the measure
\[
    \varphi_{t,t'}[H]=\varphi[\{t\}\times\{t'\}\times H]
\]
for every measurable $H\subseteq\mathcal{A}$, and let
\[
    \mu_t[t']=\frac{1}{\nu[t]}(\varphi_{t,t'}[\A]+\varphi_{t',t}[\A]).
\]
The absence of unmatched agents and condition
\eqref{eq: roommate accounting} imply that
$\mu_t\in\Delta(\Theta_\nu)$. Moreover, $\nu[t]\mu_t[t']=\nu[t']\mu_{t'}[t]$, so $\mu$ satisfies the consistency condition. For distinct types $t>t'$ such that $\mu_t[t']>0$, define
\[\gamma_{t,t'}[H] =\frac{\varphi_{t,t'}[H]+\varphi_{t',t}[H]}{\nu[t]\mu_t[t']} \quad\text{and}\quad \gamma_{t',t}[H] = \gamma_{t,t'}[\rho(H)]
\]
for every measurable $H\subseteq\A$. For each $t$ such that
$\mu_t[t]>0$, define
\[\gamma_{t,t}[H] = \frac{\varphi_{t,t}[H]+\varphi_{t,t}[\rho(H)]}{\nu[t]\mu_t[t]}.
\]
For zero-mass pairs, choose the agreement distributions jointly so that exchangeability holds. The resulting agreement profile is therefore exchangeable.

Finally, we show that $(\mu,\gamma)$ is complete-information stable. First, it satisfies internal stability. This is because $(t,t',\alpha)\in\supp(\varphi)$ implies $\alpha\in\mathbb{C}(t,t')$: for distinct types, every contract in the numerator is a correlated equilibrium written in the canonical orientation; for same-type matches, both $\varphi_{t,t}$ and its role reversal are supported on $\CE_{t,t}$.

It remains to verify external stability. Note that every current position in the induced outcome corresponds to a current position in $\varphi$ with the same expected utility. There are two cases to consider.

Suppose the induced outcome admitted a complete-information blocking pair consisting of two distinct types. Orient the proposed agreement according to the fixed ordering of types, reversing it if necessary. The resulting contract belongs to the relevant contract set and makes both agents strictly better off than their corresponding positions in $\varphi$. It therefore generates a roommate blocking pair, contradicting the stability of $\varphi$.

Suppose two agents of the same type $t$ with current utilities $u_1$ and $u_2$ form a complete-information blocking pair through an agreement $\hat{\alpha}\in\CE_{t,t}$. The symmetrized agreement
$\bar{\alpha}=\hat{\alpha}/2+\rho(\hat{\alpha})/2$ belongs to $\CE^F_{t,t}$ and gives both agents
\[
    \mathbb{E}_{\bar{\alpha}}[u_t(a,a',t)]>\min\{u_1,u_2\}.
\]
Consequently, two type-$t$ agents drawn from the worse of the two current positions can form a roommate blocking pair under $\bar{\alpha}$, again contradicting the stability of $\varphi$. The induced outcome $(\mu,\gamma)$ is therefore complete-information stable.
\end{proof}

We say that the roommate market $\mathcal{E}$ is \textbf{acyclic} if $\succ_t$ is acyclic for each $t\in\Theta_\nu$.\footnote{A relation $\succ$ on a set $Z$ is acyclic if there is no finite sequence $\{z_1,z_2,\ldots,z_n\}$ such that $z_1\succ z_2\succ\cdots\succ z_n\succ z_1$.} Moreover, $\mathcal{E}$ is \textbf{continuous} if $\{(t,\alpha,t',\alpha',t^*)\in (\widebar{\Theta}_\nu\times C)^2\times \Theta_\nu:(t,\alpha)\succ_{t^*}(t',\alpha')\}$ is open, $\mathbb{C}$ is continuous with nonempty and compact values, and $\Theta_\nu\times C$ is closed. We use the following existence result, which is Corollary 2 of \citet{carmona2024}.

\begin{lemma}[\citealp{carmona2024}]\label{lemma: existence cl}
If $\mathcal{E}$ is an acyclic and continuous roommate market without matching externalities, then $\mathcal{E}$ admits a stable roommate matching.
\end{lemma}

It remains to verify these conditions. Because $\succ_t$ is induced by a continuous utility function $v_t$, it is acyclic and its strict-preference graph is open. Moreover, the market has no matching externalities because $v_t$ does not depend on the aggregate matching. Because $A$ is finite, $\A$ is compact. Each $\CE_{t,t'}$ is a nonempty compact polytope, while each $\CE^F_{t,t}$ is nonempty and compact because it is the intersection of $\CE_{t,t}$ with the zero set of a continuous linear function. Since $\Theta_\nu$ is finite, $\mathbb{C}$ is continuous with nonempty compact values, and $\Theta_\nu\times C$ is closed. Lemmas \ref{lemma: existence cl} and \ref{lemma: equivalent stability} therefore establish the existence of a complete-information stable outcome under population distribution $\nu$.

\subsubsection{Proof of Proposition 
\ref{prop: necessary condition for polymorphism}}
\label{proof: necessary condition for polymorphism}

For part (i), fix a complete-information stable outcome $(\mu,\gamma)$ under $\nu$ and suppose $G_\theta(\mu,\gamma)>G_\tau(\mu,\gamma)$ for some $\theta,\tau\in\Theta_\nu$. Because the aggregate material payoff of the population cannot exceed $m/2$, we must have $G_t(\mu,\gamma)<m/2$ for some $t\in\Theta_\nu$. For if not, the strict payoff inequality between $\theta$ and $\tau$ would imply that the aggregate material payoff exceeds $m/2$. Choose a parochial efficient type $t'\notin\Theta_\nu$, which is possible because $\Theta_\nu$ is finite and the class of parochial efficient preferences is infinite. For some $\varepsilon\in(0,\bar{\varepsilon})$, let $\tilde{\nu}=(1-\varepsilon)\nu+\varepsilon\delta_{t'}$. Construct an outcome $(\tilde{\mu},\tilde{\gamma})$ under $\tilde{\nu}$ by preserving the matching and play of all incumbent types, while matching type-$t'$ agents only with themselves and having them play efficient agreements. Because the masses of all incumbent types are scaled by the same factor, $\tilde{\mu}$ satisfies consistency. Moreover, the incumbents' current positions are unchanged, while type-$t'$ agents attain their highest feasible utility in same-type matches. Thus, $(\tilde{\mu},\tilde{\gamma})$ is complete-information stable. Yet, $G_{t'}(\tilde{\mu},\tilde{\gamma})=m/2>
G_t(\tilde{\mu},\tilde{\gamma})$, contradicting the local evolutionary dominance of $\nu$. This proves part (i).

For part (ii), suppose $\mu_t[t']>0$ and some $\alpha\in\supp(\gamma_{t,t'})$ is inefficient. Expected joint material payoff is continuous in the agreement and strictly below $m$ at $\alpha$. It is therefore strictly below $m$ in a neighborhood of $\alpha$, which has positive probability because $\alpha\in\supp(\gamma_{t,t'})$. Aggregate material payoff is consequently strictly below $m/2$. By part (i), all incumbent types then earn strictly less than $m/2$. The parochial efficient entrant constructed above again yields a contradiction. Hence, every agreement played in a positive-mass match is efficient. Part (i) and the aggregate payoff identity then imply that, in every complete-information stable outcome under $\nu$, $G_t(\mu,\gamma)=m/2$ for every $t\in\Theta_\nu$.

We next show that every positive-mass cross-type match divides material payoff equally on average. For $\mu_t[t']>0$, let
\[
h_{t,t'} = \int_{\A} \mathbb{E}_{\alpha}[\pi(a,a')] \mathop{d\gamma_{t,t'}[\alpha]}.
\]
Efficiency and exchangeability then imply $h_{t,t'}+h_{t',t}=m$. In particular, $h_{t,t}=m/2$.

We now show that $h_{t,t'}=m/2$ even for cross-type matches as well. Suppose, toward a contradiction, that $h_{t,t'}>m/2$ for some distinct $t,t'\in\Theta_\nu$. Type $t'$ then receives less than $m/2$ in these matches. Since its average material payoff is $m/2$, it must receive more than $m/2$ in matches with some other type. Repeating this argument and using the finiteness of $\Theta_\nu$, we obtain distinct types $t_1,\ldots,t_k$ such that $\mu_{t_i}[t_{i+1}]>0$ and $h_{t_i,t_{i+1}}>m/2$ for every $i=1,\ldots,k$, with indices interpreted modulo $k$ so that $t_{k+1}=t_1$. We consider two cases.

Suppose first that $k$ is even. For $\zeta>0$ sufficiently small, define a new $\mu'$ by increasing the mass of matches between $t_i$ and $t_{i+1}$ by $\zeta$ when $i$ is odd and decreasing it by $\zeta$ when $i$ is even, leaving all other match masses and the agreement profile unchanged.\footnote{More specifically, for each odd $i$, add $\zeta/\nu[t_i]$ to $\mu_{t_i}[t_{i+1}]$ and subtract it from $\mu_{t_i}[t_{i-1}]$; for each even $i$, reverse these adjustments. Leave all other components unchanged.} For each type, the increase in matches with one neighboring type is exactly offset by the decrease in matches with the other. Thus, $\mu'$ remains consistent and preserves every type's total mass. For sufficiently small $\zeta$, every cycle edge remains positive. Hence, the set of current positions is unchanged, and $(\mu',\gamma)$ remains complete-information stable. However, the resulting outcome satisfies $G_{t_i}(\mu',\gamma)>G_{t_{i+1}}(\mu',\gamma)$ for all odd $i$, contradicting part (i).

Suppose next that $k$ is odd. Relabel the cycle so that $h_{t_1,t_{2}}\geq h_{t_i,t_{i+1}}$ for all $i=1,\ldots,k$. For $\varepsilon\in(0,\bar{\varepsilon})$ sufficiently small, consider $\tilde{\nu}=(1-\varepsilon)\nu+\varepsilon\delta_{t_1}$. Scale every original match mass by $1-\varepsilon$. Along the cycle, additionally increase the mass of matches between $t_i$ and $t_{i+1}$ by $\varepsilon/2$ when $i$ is odd and decrease it by $\varepsilon/2$ when $i$ is even.\footnote{Begin with the scaled profile $\tilde{\mu}_t[\cdot]=(1-\varepsilon)\nu[t]\mu_t[\cdot]/\tilde{\nu}[t]$. Add $\varepsilon/(2\tilde{\nu}[t_1])$ to both $\tilde{\mu}_{t_1}[t_2]$ and $\tilde{\mu}_{t_1}[t_k]$. For each odd $i\neq 1$, add $\varepsilon/(2\tilde{\nu}[t_i])$ to $\tilde{\mu}_{t_i}[t_{i+1}]$ and subtract it from $\tilde{\mu}_{t_i}[t_{i-1}]$; for each even $i$, reverse these adjustments. Leave all other components unchanged.} Because $k$ is odd, the masses of matches between $t_1$ and its two neighboring types each increase by $\varepsilon/2$, together absorbing the additional mass $\varepsilon$ of type $t_1$. For every other type, the increase in matches with one neighbor is exactly offset by the decrease in matches with the other. The resulting matching profile $\tilde{\mu}$ is therefore consistent. For sufficiently small $\varepsilon$, its support is unchanged, so $(\tilde{\mu},\gamma)$ is complete-information stable. However, it is easy to verify that $G_{t_1}(\tilde{\mu},\gamma)\geq m/2>G_{t_{i}}(\tilde{\mu},\gamma)$ for every even $i$, contradicting the assumption that $\nu$ is locally evolutionarily dominant. We conclude that $h_{t,t'}=m/2$ for every positive-mass cross-type match.

Finally, fix distinct $t,t'\in\Theta_\nu$ and $\alpha\in\supp(\gamma_{t,t'})$. Replace $\gamma_{t,t'}$ by the point mass on $\alpha$ and $\gamma_{t',t}$ by the point mass on $\rho(\alpha)$, leaving all other components unchanged. The resulting outcome remains internally stable, and its set of current positions is a subset of that under $(\mu,\gamma)$; hence, it also remains externally stable. Applying the preceding conclusion to this outcome gives
\[ \mathbb{E}_{\alpha}[\pi(a,a')] = \mathbb{E}_{\rho(\alpha)}[\pi(a,a')] = \frac{m}{2}.\]
This proves part (ii).

\end{appendix}

\end{document}